\documentclass{ws-nl}
\usepackage{graphicx}
\usepackage{amssymb}
\usepackage[american]{babel}
\usepackage{epsfig}
\usepackage{float}
\usepackage{subfigure}
\usepackage{placeins}
\usepackage{tabularx}

\usepackage{multicol}
\usepackage[unicode=true,
 bookmarks=true,bookmarksnumbered=false,bookmarksopen=false,
 breaklinks=false,pdfborder={0 0 0},backref=false,colorlinks=true]
 {hyperref}
\hypersetup{pdftitle={Large-scale first principles configuration interaction calculations of optical absorption in boron clusters,- Ravindra Shinde and Alok Shukla},
  pdfauthor={Ravindra Shinde, Alok Shukla},
  pdfsubject={Computational Condensed Matter},
  unicode=false,pdftoolbar=true,pdfmenubar=true,pdffitwindow=true,pdfstartview={FitH},pdfnewwindow=true,pdfcreator={Latex2e},linkcolor=red,citecolor=blue,urlcolor=magenta}
\usepackage[superscript,nocompress]{cite}
\usepackage{hypcap}
\begin{document}

\catchline{\#}{\#}{2012}{}{}
\markboth{Ravindra Shinde and Alok Shukla}{Large scale first principles configuration interaction calculations of optical absorption in boron clusters}

\title{LARGE-SCALE FIRST PRINCIPLES CONFIGURATION INTERACTION CALCULATIONS OF OPTICAL ABSORPTION IN BORON CLUSTERS}

\author{\href{http://home.iitb.ac.in/~ravindra.shinde}{RAVINDRA SHINDE}}

\address{{Physics Department, Indian Institute of Technology Bombay, \\ 
Mumbai, Maharashtra 400076, INDIA.}\\
\email{ravindra.shinde@iitb.ac.in}
}

\author{\href{http://www.phy.iitb.ac.in/doku/doku.php/faculty/shukla/home}{ALOK SHUKLA}}
\address{Physics Department, Indian Institute of Technology Bombay, \\
Mumbai, Maharashtra 400076, INDIA.\\
\email{shukla@phy.iitb.ac.in}}

\maketitle

\begin{history}
\received{\today}
\end{history}

\begin{abstract}
We have performed systematic large-scale all-electron correlated calculations
on boron clusters B$_{n}$(n=2--5), to study their linear optical
absorption spectra. Several possible isomers of each cluster were
considered, and their geometries were optimized at the coupled-cluster
singles doubles (CCSD) level of theory. Using the optimized ground-state
geometries, excited states of different clusters were computed using
the multi-reference singles-doubles configuration-interaction (MRSDCI)
approach, which includes electron correlation effects at a sophisticated
level. These CI wave functions were used to compute the transition
dipole matrix elements connecting the ground and various excited states
of different clusters, eventually leading to their linear absorption
spectra. The convergence of our results with respect to the basis
sets, and the size of the CI expansion were carefully examined. The
contribution of configurations to many body wavefunction of various
excited states suggests that the excitations involved are collective,
plasmonic type.
\end{abstract}

\keywords{photo-absorption, cluster, configuration interaction, boron, optical, MRSDCI}

\begin{multicols}{2}
\section{\label{sec:intro}INTRODUCTION}
The area of cluster science has witnessed an enormous progress in
terms of both the experimental and theoretical investigations of clusters
of various atoms over last few decades.\cite{alonso_book,clustnano_book,julius_book,deheer_rmp}
Ranging from small clusters having a few atoms, to nanotubes, nanosheets,
fullerenes \emph{etc.}, clusters have proven their usefulness in the
fast emerging field of nanotechnology.\cite{potential_boron} Because
of the finiteness of the size, properties of clusters can be greatly
tuned, and, therefore they are more amenable to nanoengineering than
their bulk counterpart.\cite{tunable_icosa} The evolution of clusters
towards the bulk with the increasing number of atoms, and the underlying
mechanism, is a research topic of great contemporary interest.\cite{fullerene_form1,fullerene_form2}

At the present time, boron clusters are attracting great attention
because of their novel properties, and potential applications in nanotechnology
and hydrogen storage related capabilities.\cite{boron_age2004,zhao_prl2005,cabria_alonso,shelvin_guo}
Boron atom, having $s^{2}p^{1}$ valence electronic configuration,
has short covalent radius and tends to form strong directional bonds
producing clusters of covalent nature. Because of this strong covalent
bonding, it has hardness close to that of diamond. The ability of
boron to form structures of any size due to catenation is only comparable
to its neighbor carbon.\cite{boron_age2004} Planar boron clusters
exhibit aromaticity\cite{borozene_sahu} due to the presence of itinerant
$\pi$ electrons, and some of them are analogous to aromatic hydrocarbons.\cite{kiran_wang}
Boron fullerenes, boron sheets and single-sheet boron nitride---a
graphene analogue---are the other examples of boron-based clusters.

As far as the studies of boron-based clusters are concerned, small
ionic boron clusters B$_{n}^{+}$ (n $\leqslant$ 20) were experimentally
studied by Hanley, Whitten and Anderson.\cite{hanley_whitten} Wang
and coworkers have reported joint theoretical and experimental studies
of the electronic structure of bare boron wheels, rings, tubes and
large quasi-planar clusters.\cite{wheel_wang,ring_wang,tube_wang,kiran_wang}
Using the photoelectron spectroscopy, they predicted that tubular
B$_{20}$ can act as the smallest boron single walled nanotube. Transition
metal-centered boron ionic ring clusters were studied by Constantin
\emph{et. al.} \cite{ring_wang}, in a photo-electron spectroscopy
experiment, supported by first-principles calculations. The abundance
spectrum of boron clusters generated by laser ablation of hexagonal
boron nitride was studied by time of flight measurements performed
by La Placa, Roland and Wynne.\cite{la_placa} They also postulated
the existence of B$_{36}$N$_{24}$ cluster having a structure similar
to that of C$_{60}$ fullerene. Lauret \emph{et al.}\cite{prl_lauret}
probed the optical transitions in single walled boron nitride nanotubes
by means of optical absorption spectroscopy.

Larger pure boron clusters have also been investigated extensively.
Cage-like structure of B$_{80}$---similar to C$_{60}$ fullerene---has
been proposed theoretically.\cite{prl_szwacky} A density functional
theory (DFT) study of pure boron sheets and nanotubes was carried
out by Cabria, Lopez and Alonso to explore their potential hydrogen
storage materials.\cite{cabria_alonso} Chacko, Kanhere and Boustani
investigated different equilibrium geometries of B$_{24}$ cluster
using Born-Oppenheimer molecular dynamics within the framework of
DFT.\cite{chacko_kanhere} Abdurahman \emph{et al.}\cite{ayjamal}
studied the ladder-like planar boron chains B$_{n}$ ($n$=4-14),
and computed their static dipole polarizabilities using the \emph{ab
initio} CI method. Johansson discussed strong toroidal ring currents
in B$_{20}$ and other toroidal boron clusters.\cite{toroid_current}
Double aromaticity was proposed in toroidal boron clusters B$_{2n}$
(n = 6,14) by Bean and Fowler.\cite{aromatic_toroid}

Regarding the smaller sized boron clusters, an early theoretical study
of boron dimer was carried out by Langhoff and Bauschlicher,\cite{langhoff}
who performed an extensive calculations using the complete-active-space
self-consistent-field (CASSCF) multireference configuration interaction
(MRCI) with a large basis set. A similar study was carried out by
Bruna and Wright\cite{bruna_wright} for the excited states of B$_{2}$,
and by Howard and Ray\cite{howard_ray} using the many-body perturbation
theory. A systematic geometry and electronic structure calculations
of bare boron clusters was reported by Boustani.\cite{boustani_prb97}
He performed all-electron calculations at the SDCI level, but the
contracted Gaussian basis sets used were small. Niu, Rao and Jena,\cite{rao_jena}
using DFT and quantum chemical methods, presented an account of electronic
structures of neutral and charged boron clusters. In their study on
small clusters, M\"{o}ller-Plesset perturbation theory of fourth order
(MP4) was used to account for the electron correlation effects. More
recently, Ati\c{s}, \"{O}zdogan, and G\"{u}ven\c{c} investigated structure and
energetics of boron clusters using the DFT.\cite{turkish_boron} Aromaticity
in planar boron clusters was addressed by Aihara, Kanno and Ishida.\cite{ishida_jacs}

In spite of many theoretical studies of boron clusters of various
shapes and sizes, very little experimental information about their
ground and excited states is available. Conventional mass spectrometry
can distinguish between different clusters only according to their
mass, but not according to their geometry. One has to rely on other
theoretical or experimental data to be able to differentiate one isomer
from another. For example, using first principles calculations of
vibrionic fine structure in C$_{20}^{-}$, and comparing it with experimentally
available data, Saito and Y. Miyamoto\cite{saito_fullerene} identified
the cage and bowl structures. Optical absorption spectroscopy, coupled
with extensive theoretical calculations of the optical absorption
spectra, can be used to distinguish between distinct isomers of clusters
produced experimentally, because normally optical absorption spectra
are sensitive to the geometries of the clusters. The optical absorption
of alkali metal clusters has been extensively studied both experimentally
and theoretically.\cite{na_opt_kappes_jcp,na_dehaar_prl,na_opt_bethe_salp,li_boustani_prb,li_na_tddft}
However, a very few such studies exist for the case of boron clusters.
Marques and Botti\cite{marques_botti} calculated optical absorption
on different B$_{20}$ isomers using time-dependent (TD) DFT. Boron
fullerenes such as B$_{38}$, B$_{44}$, B$_{80}$ and B$_{92}$ were
also studied by Botti and coworkers\cite{marques_fullerene} using
the same technique. However, to the best of out knowledge, there are
no experimental and theoretical study of optical absorption on other
bare boron clusters, particularly the smaller ones. It is with the
aim of filling this void that we undertake a systematic study of the
optical absorption in small boron clusters B$_{n}$(n=2--5), employing
the MRSDCI method, and high-quality Gaussian basis functions. We perform
careful geometry optimization for each possible isomer, and compute
the optical absorption spectra of various structures. We also analyze
the many-body wave functions of various excited states contributing
to the peaks in the computed spectra, and conclude that most of the
excitations are collective in nature, signalling the presence of plasmons.

The remainder of this paper is organized as follows. Next section
describes the theoretical and computational details of the work, followed
by section \ref{sec:results} in which our results are presented and
discussed. In section  \ref{sec:conclusions} we present our conclusions
and discuss possibilities for future work. Detailed information about
various excited states contributing to optical absorption is presented
in the Appendix. 

\section{\label{sec:theory}THEORETICAL AND COMPUTATIONAL DETAILS}

\begin{figure*}
\begin{center}
\subfigure[B$_{2}$, D$_{\infty h}$, $^{3}\Sigma_{g}^{-}$ \label{fig:b2geom}]{\psfig{file=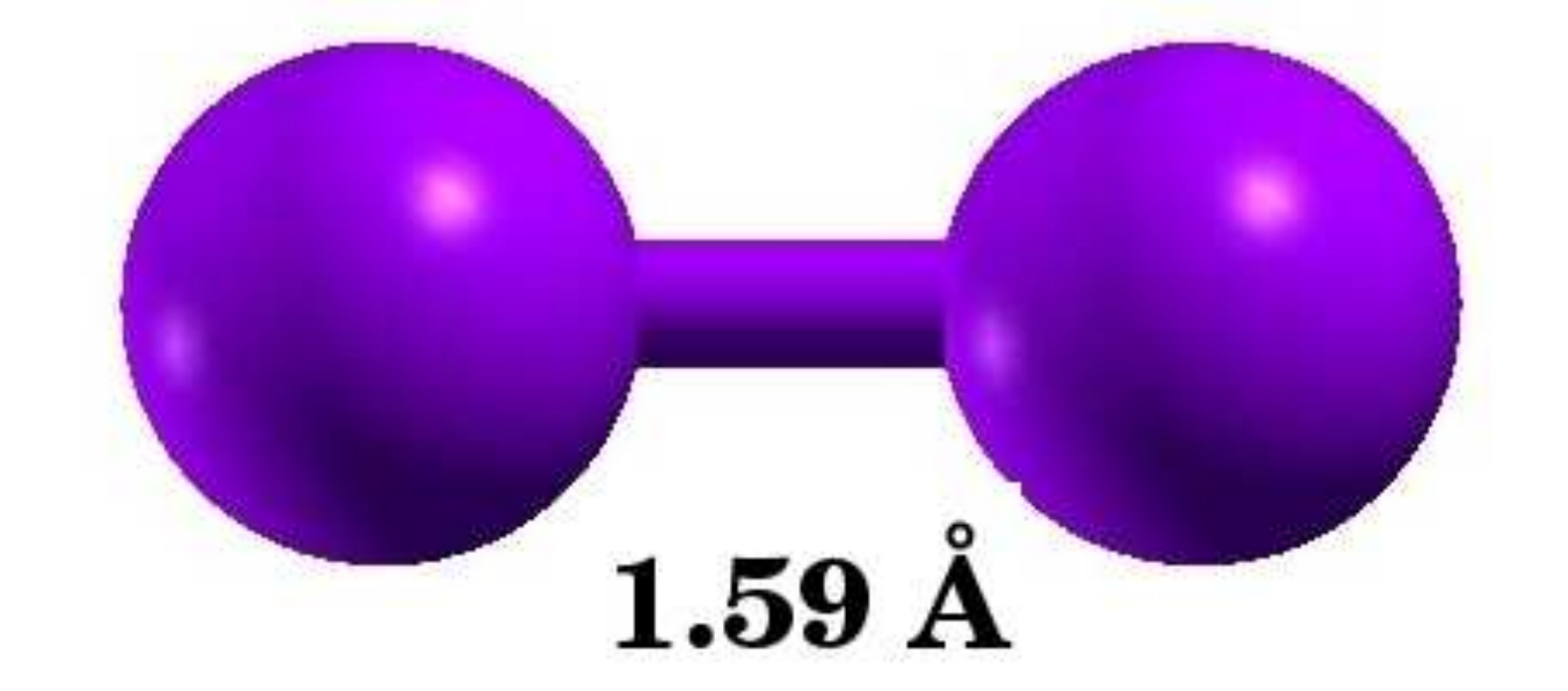,width=3.1cm}}
\hspace{0.4cm} \subfigure[B$_{3}$, D$_{3h}$, $^{2}$A$_{1}^{'}$ \label{fig:b3trgeom} ]{\psfig{file=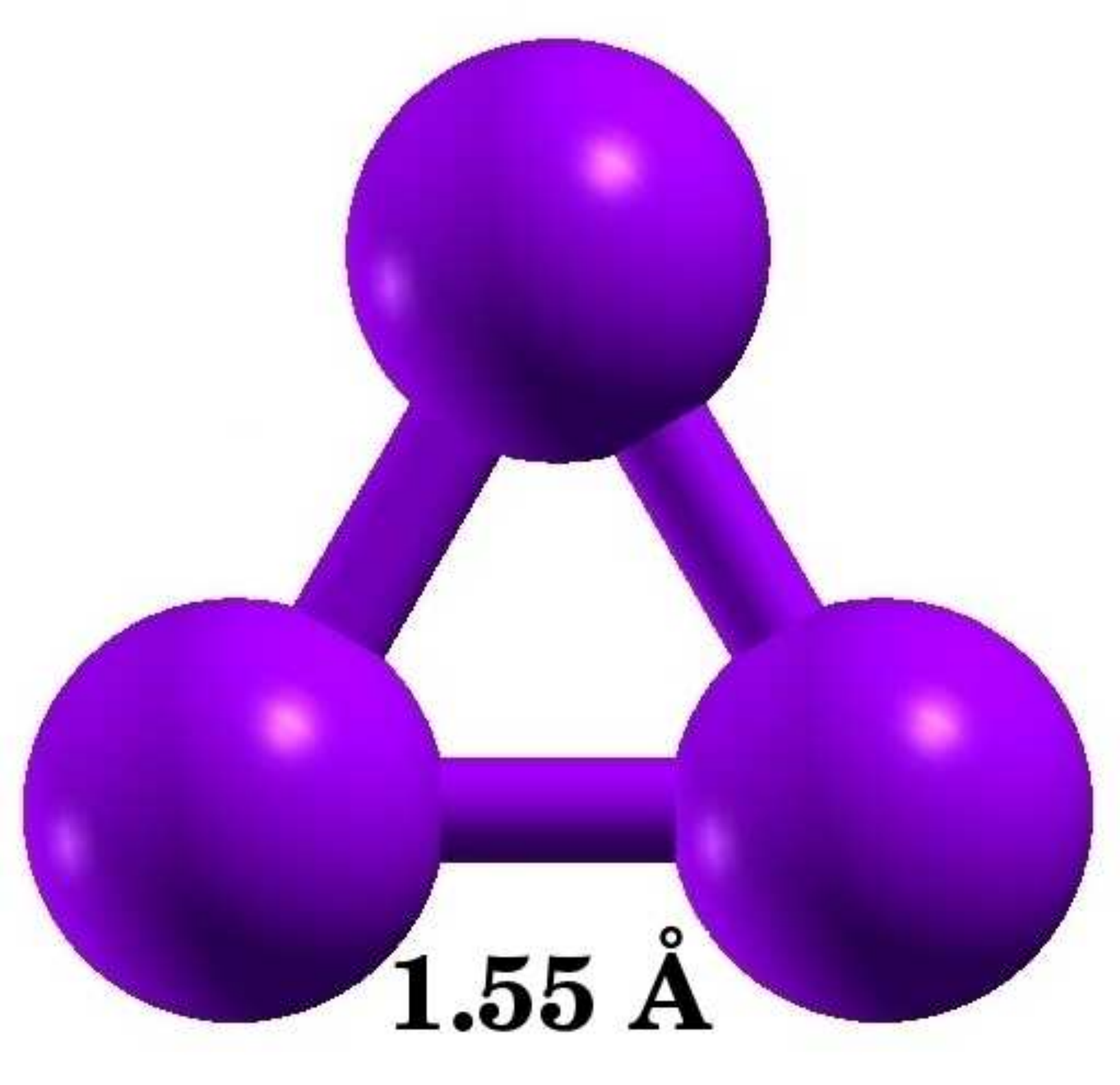,width=2.6cm}}
\hspace{0.4cm} \subfigure[B$_{3}$, D$_{\infty h}$, $^{2}\Sigma_{g}^{-}$ \label{fig:b3lingeom}]{\psfig{file=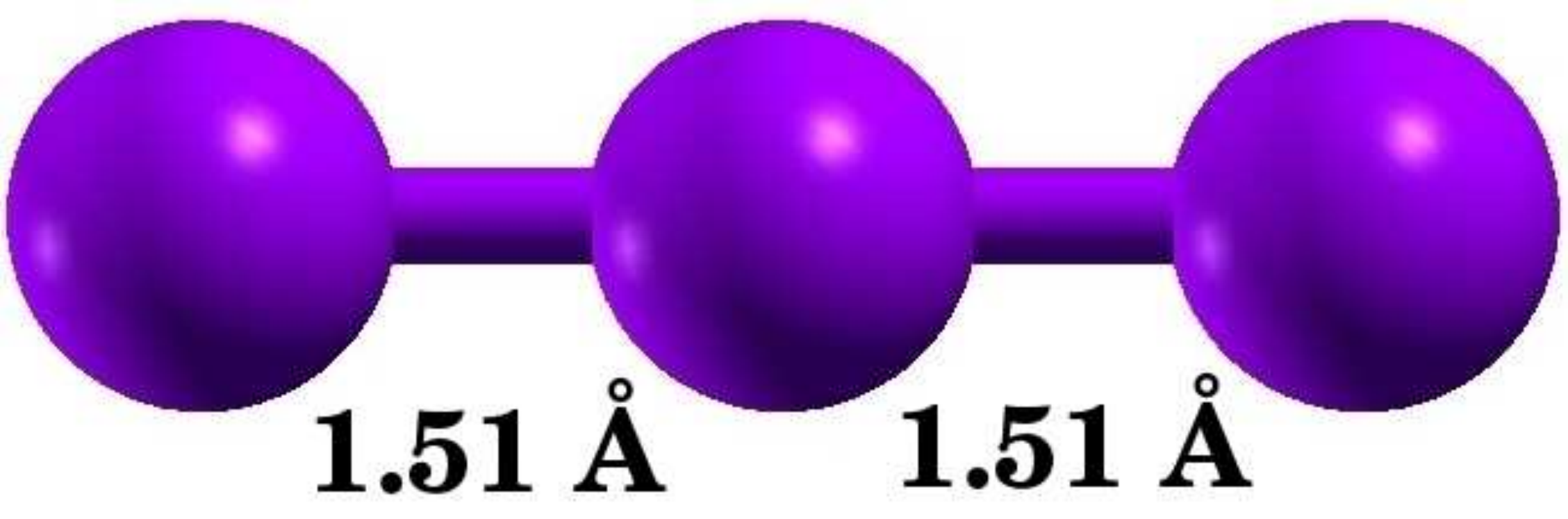,scale=0.17}}  
\hspace{0.4cm} \subfigure[B$_{4}$, D$_{2h}$, $^{1}$A$_{g}$ \label{fig:b4rhogeom}]{\psfig{file=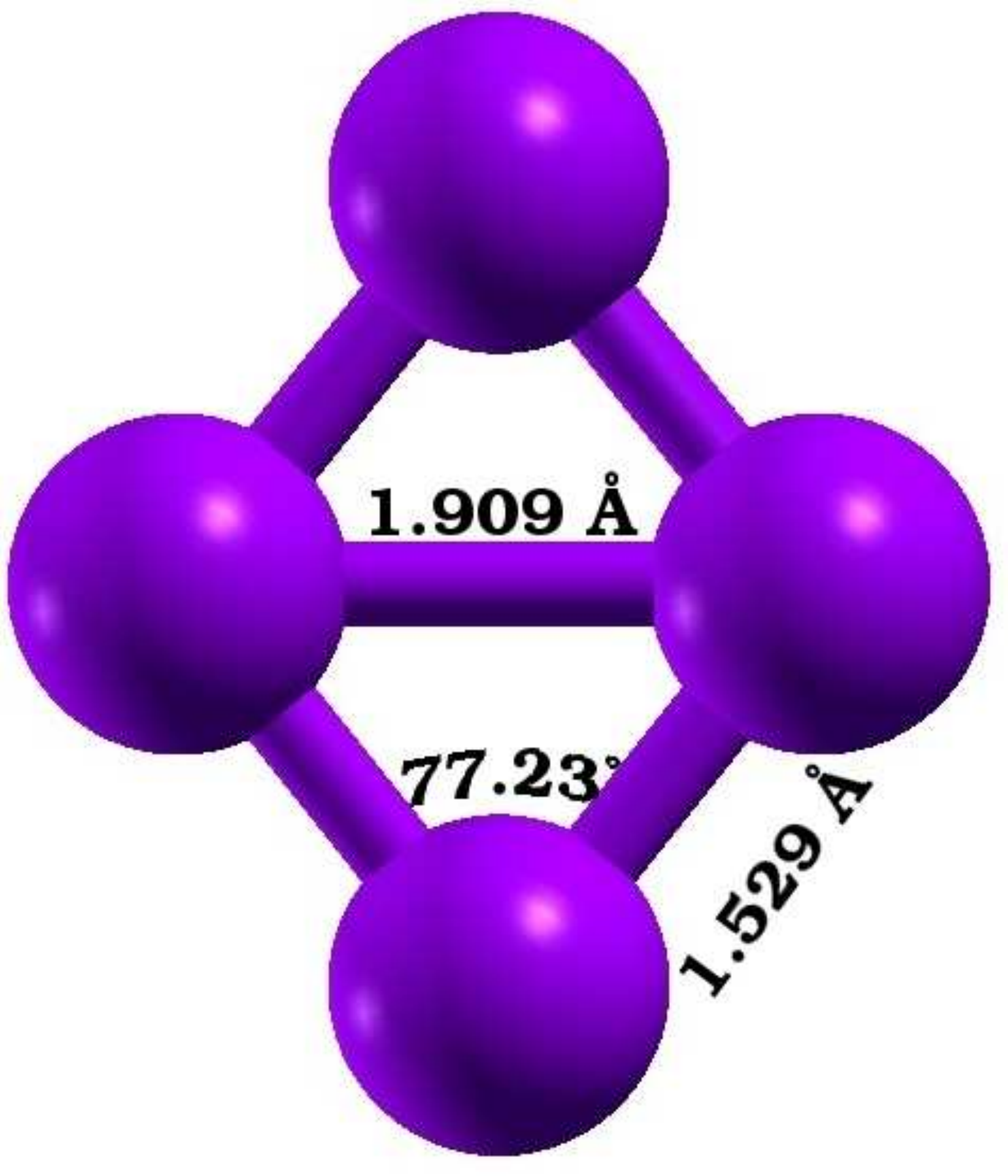,width=2.9cm}} 
 \hspace{0.4cm}\subfigure[B$_{4}$, D$_{4h}$, $^{1}$A$_{1g}$ \label{fig:b4sqrgeom}]{\psfig{file=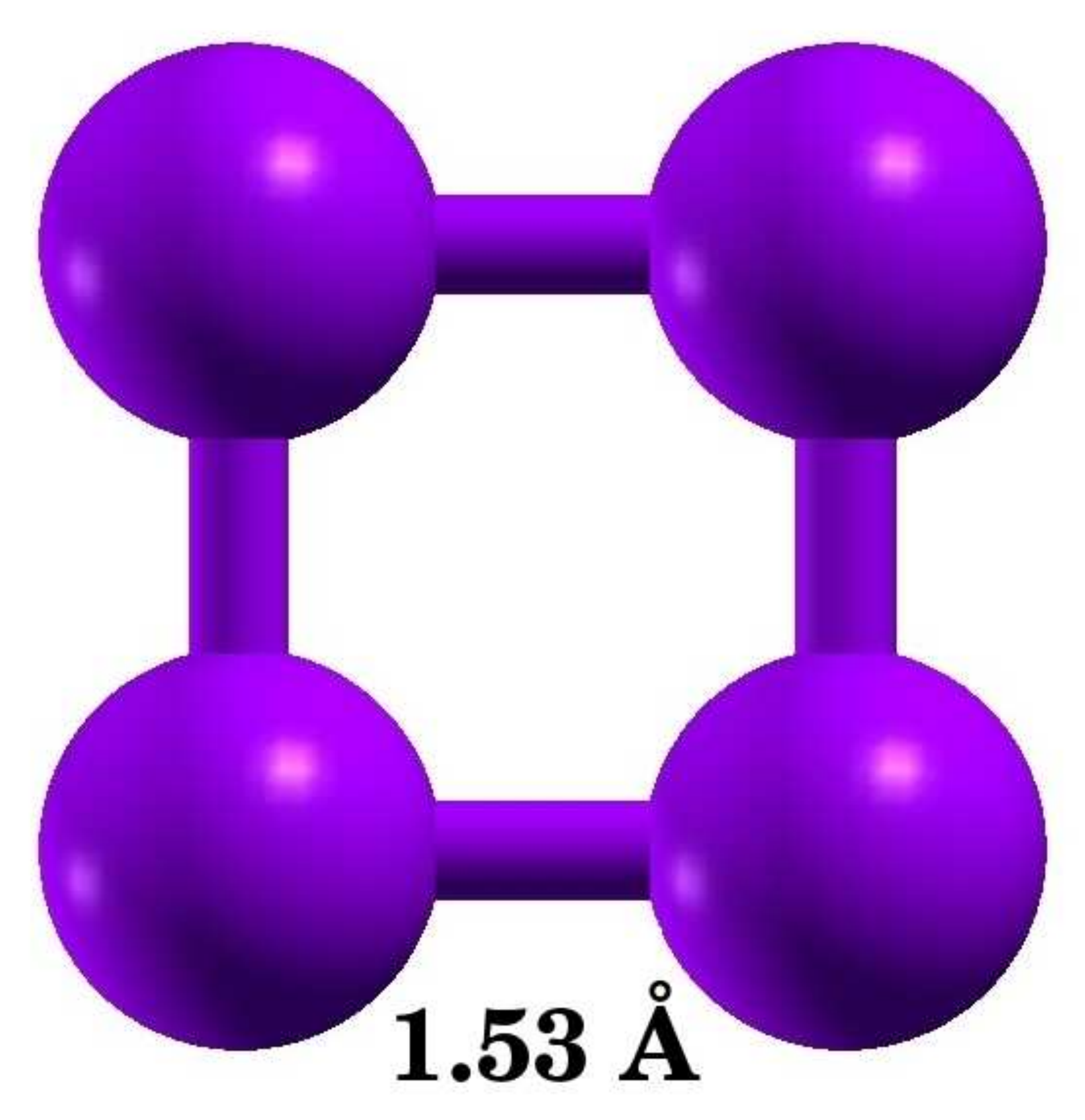,width=2.7cm}} 
 \hspace{0.4cm}\subfigure[B$_{4}$, D$_{\infty h}$, $^{1}\Sigma_{g}^{-}$ \label{fig:b4lingeom}]{\psfig{file=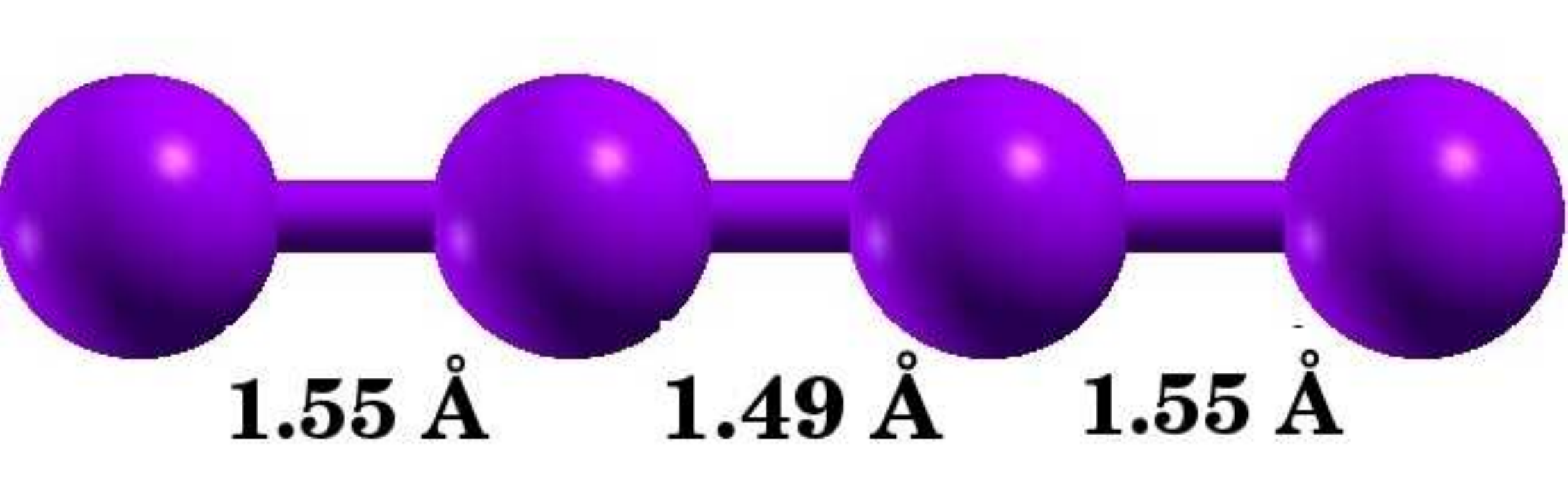,scale=0.23}} 
               \subfigure[B$_{4}$, C$_{2v}$, $^{1}$A$_{1}$ \label{fig:b4tetgeom}]{\psfig{file=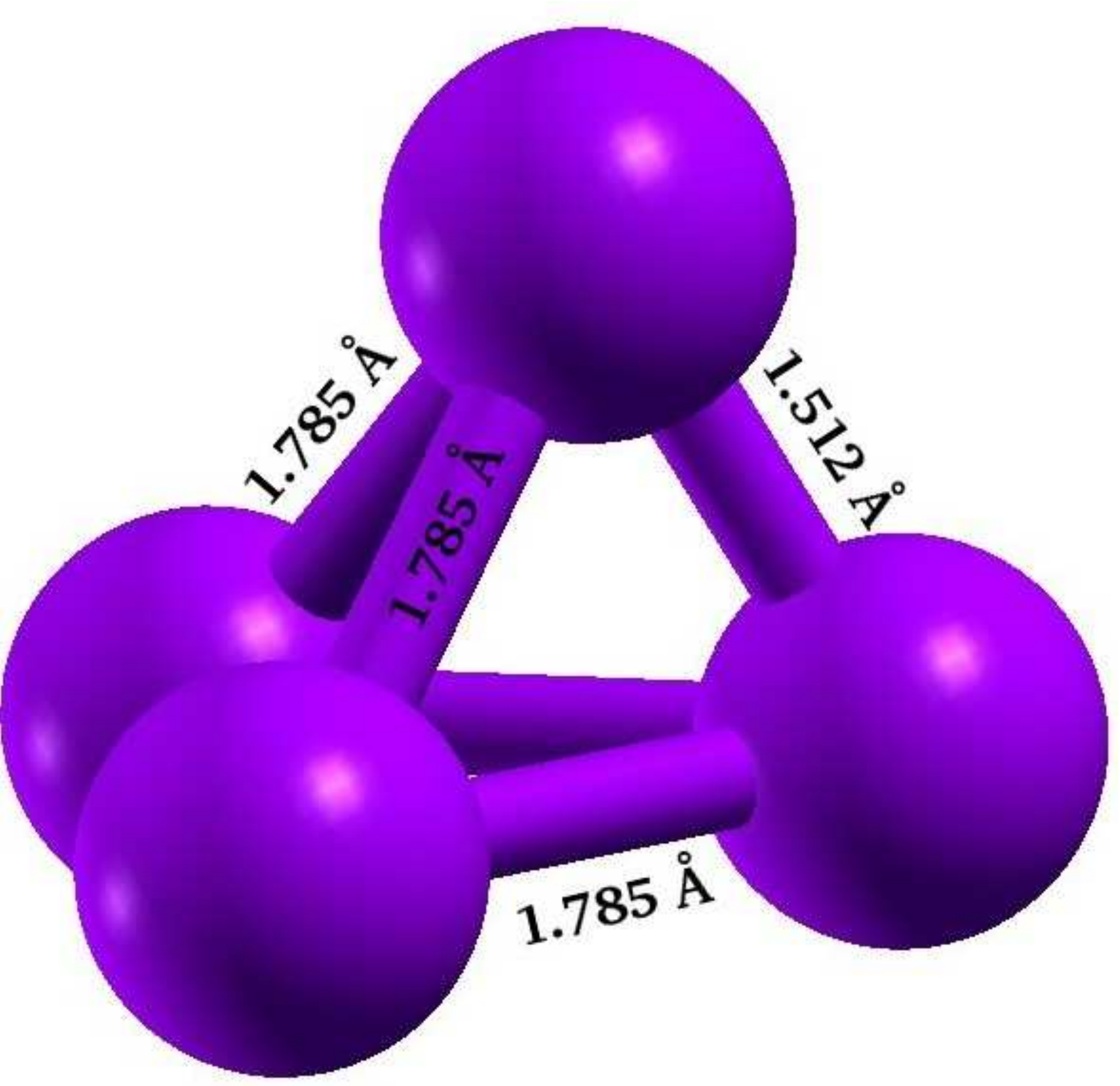,width=2.9cm}}
 \hspace{0.4cm}\subfigure[B$_{5}$, C$_{2v}$, $^{2}$B$_{2}$ \label{fig:b5pengeom}]{\psfig{file=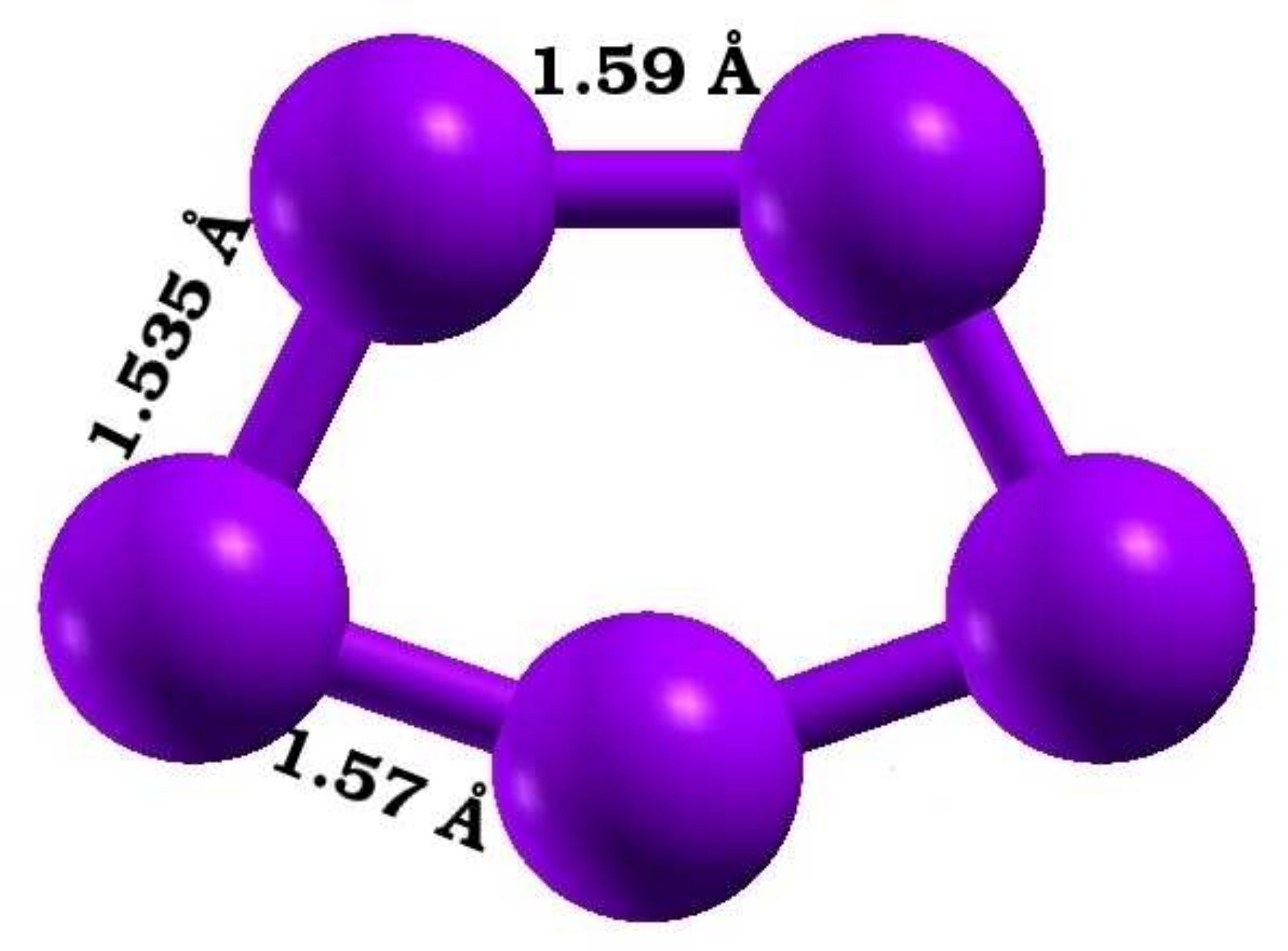,width=4.5cm}}
 \hspace{0.4cm}\subfigure[B$_{5}$, C$_{s}$, $^{2}$A \label{fig:b5tetgeom}]{\psfig{file=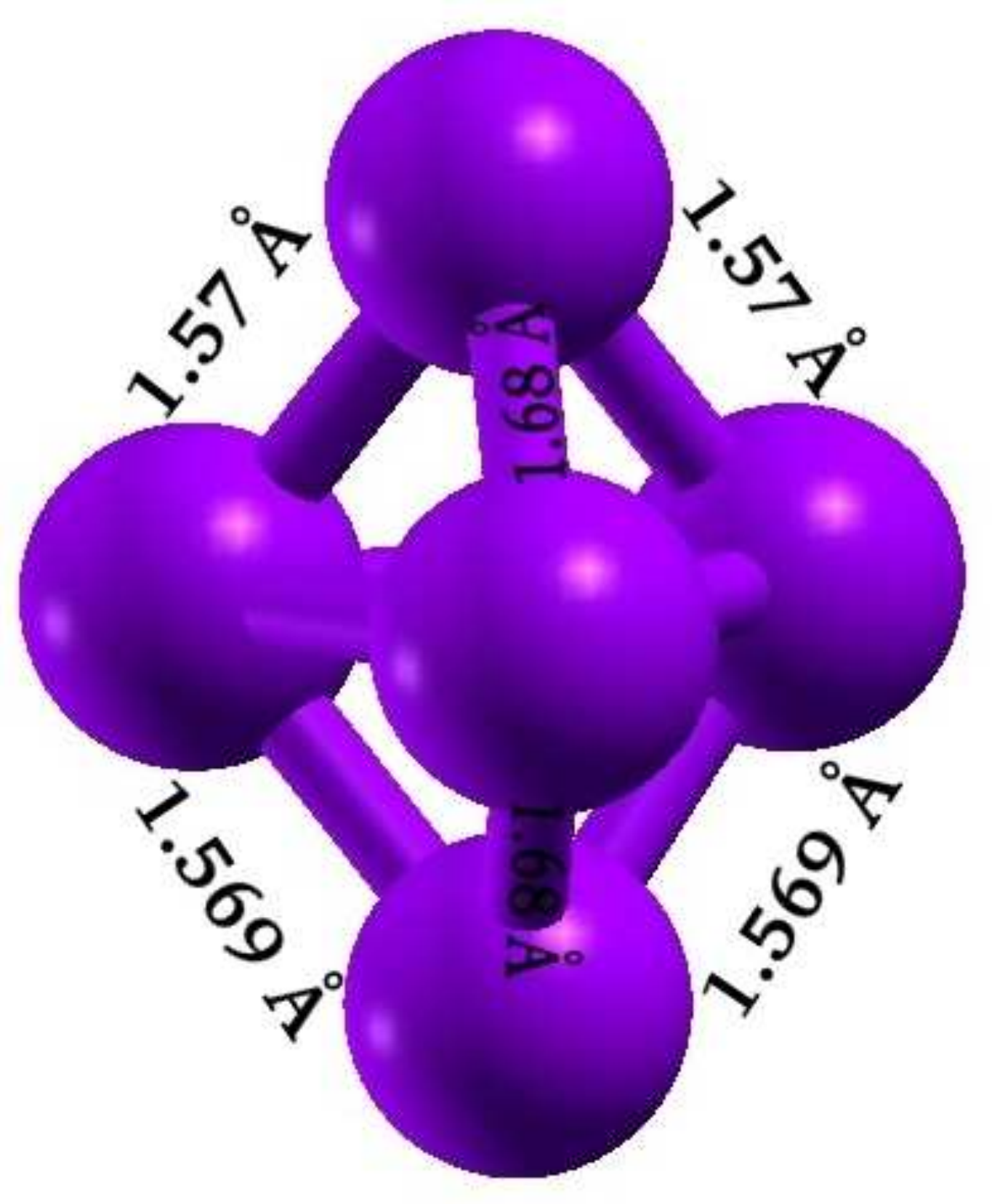,width=2.8cm}}
 \caption{(Color online) Geometry optimized structures of boron clusters with
  point group symmetry and the electronic ground state at the configuration
  interaction level.}
 \label{fig:geometries} 
\end{center}
 \end{figure*}

The geometry optimization of various isomers was done using the size-consistent
coupled-cluster singles doubles (CCSD) based analytical gradient approach,
as implemented in the package \texttt{\textsc{gamess-us}}.\cite{gamess}
For the purpose, we used the 6-311G(d,p) basis set included in the
program library,\cite{gamess} which is known to be well-suited for
this task.  The process of optimization was initiated by using the
geometries reported by Ati\c{s} \emph{et al.}\cite{turkish_boron},
based upon first principles DFT based calculations. For some simple
geometries such as B$_{2}$, B$_{3}$ (D$_{3h}$ symmetry), the optimization
was carried out manually, by performing the MRSDCI calculations at
different geometries, and locating the energy minima. Figure \ref{fig:geometries}
shows the final optimized geometries of the isomers studied in this
paper.

Once the ground state geometries of various isomers were determined,
the correlated calculations were performed to calculate their low-lying
excited states using the multi-reference singles-doubles configuration-interaction
(MRSDCI) approach as implemented in the computer program MELD.\cite{meld}
MRSDCI approach is a well-established quantum-chemical approach in
which one considers singly- and doubly-excited configurations from
a number of reference configurations, leading to a good treatment
of electron correlations, both for the ground, and the excited states,
in the same calculation. Using the ground- and excited-state wave
functions obtained from the MRSDCI calculations, electric dipole matrix
elements are computed and subsequently utilized to compute the linear
absorption spectrum assuming a Lorentzian line shape. By analyzing
the wave functions of the excited states contributing to the peaks
of the computed spectrum obtained from a given calculation, bigger
MRSDCI calculations were performed by including a larger number of
reference states. The choice of the reference states to be included
in a given calculation was based upon the magnitude of the corresponding
coefficients in the CI wave function of the excited state (or states)
contributing to a peak in the spectrum. This procedure was repeated
until the computed spectrum converged within an acceptable tolerance,
and all the configurations contributing significantly to various excited
states were included in the list of the reference states. In the past,
we have used such an iterative MRSDCI approach on a number of conjugated
polymers to perform large-scale correlated calculations of their linear
and nonlinear optical properties.\cite{mrsd_jcp_09,mrsd_prb_07,mrsd_prb_05,mrsd_prb_02}

The number of molecular orbitals, and thus the size of the CI expansion,
increases rapidly with the increasing number of atoms in the clusters.
Such a proliferation in the size of calculations can essentially make
high-quality MRSDCI calculations impossible even for clusters of the
sizes discussed in this work. Therefore, wherever possible, we have
used the point-group symmetries corresponding to $D_{2h}$, and its
subgroups, at all levels of calculations to reduce the size of the
CI expansions. During the MRSDCI calculatiosn, the frozen-core approximation
was employed, \emph{i.e.}, while constructing the CI expansion, no
virtual excitations from the $1s$ core electrons of the boron atoms
of the cluster were considered. Similarly, excitations into very high
energy virtual orbitals were not considered with the purpose of keeping
the calculations manageable. The impact of both the frozen-core approximation,
and the deletion of high-energy virtual orbitals, along with the influence
of the choice of the basis sets on our calculations will be examined
carefully in the next section.

\section{\label{sec:results}RESULTS AND DISCUSSION}

In this section, first we discuss the convergence of our calculations
with respect to various approximations and truncation schemes. Thereafter,
we present and discuss the results of our calculations for various
clusters.

\subsection{Convergence of Calculations}

Here, we carefully examine the convergence of the calculated absorption
spectra with respect to the size and quality of the basis set, along
with various truncation schemes in the CI calculations.

\subsubsection{Choice of the basis set}

In general, the results of electronic structure calculations depend
upon the quality and the size of the basis set employed. While several
contracted Gaussian basis functions have been devised which can deliver
high-quality results on various quantities such as the total energy,
correlation energy, and the static polarizabilities of molecules,
to the best of our knowledge the basis set dependence of linear optical
absorption has not been explored. Since boron shows strong covalent
bondings, the basis set used for calculations should have diffuse
Gaussian contractions. Therefore, to explore the basis set dependence
of computed spectra, we used several basis sets\cite{emsl_bas1,emsl_bas2}
to compute the optical absorption spectrum of the smallest cluster,
\emph{i.e.}, B$_{2}$. For the purpose, we used correlation-consistent
basis sets named AUG-CC-PVTZ, DAUG-CC-PVDZ, AUG-CC-PVDZ, CC-PVDZ,
and DZP, which consist of polarization functions along with diffuse
exponents, and were designed specifically for post Hartree-Fock correlation
calculations\cite{emsl_bas1,emsl_bas2}. From the calculated spectra
presented in Fig. \ref{fig:basis} the following trends emerge: the
spectra computed by various augmented basis sets (AUG-CC-PVTZ, DAUG-CC-PVDZ,
AUG-CC-PVDZ) are in good agreement with each other in the energy range
up to 8 eV, while those obtained using the nonaugmented sets (CC-PVDZ
and DZP) disagree with them substantially, particularly in the higher
energy range. Given the fact that augmented basis sets are considered
superior for molecular calculations, we decided to perform calculations
on the all the clusters using the AUG-CC-PVDZ basis set. This is the
smallest of the augmented basis sets considered by us, and, therefore,
does not cause excessive computational burden when used for larger
clusters.

\begin{figurehere}
\centerline{\psfig{file=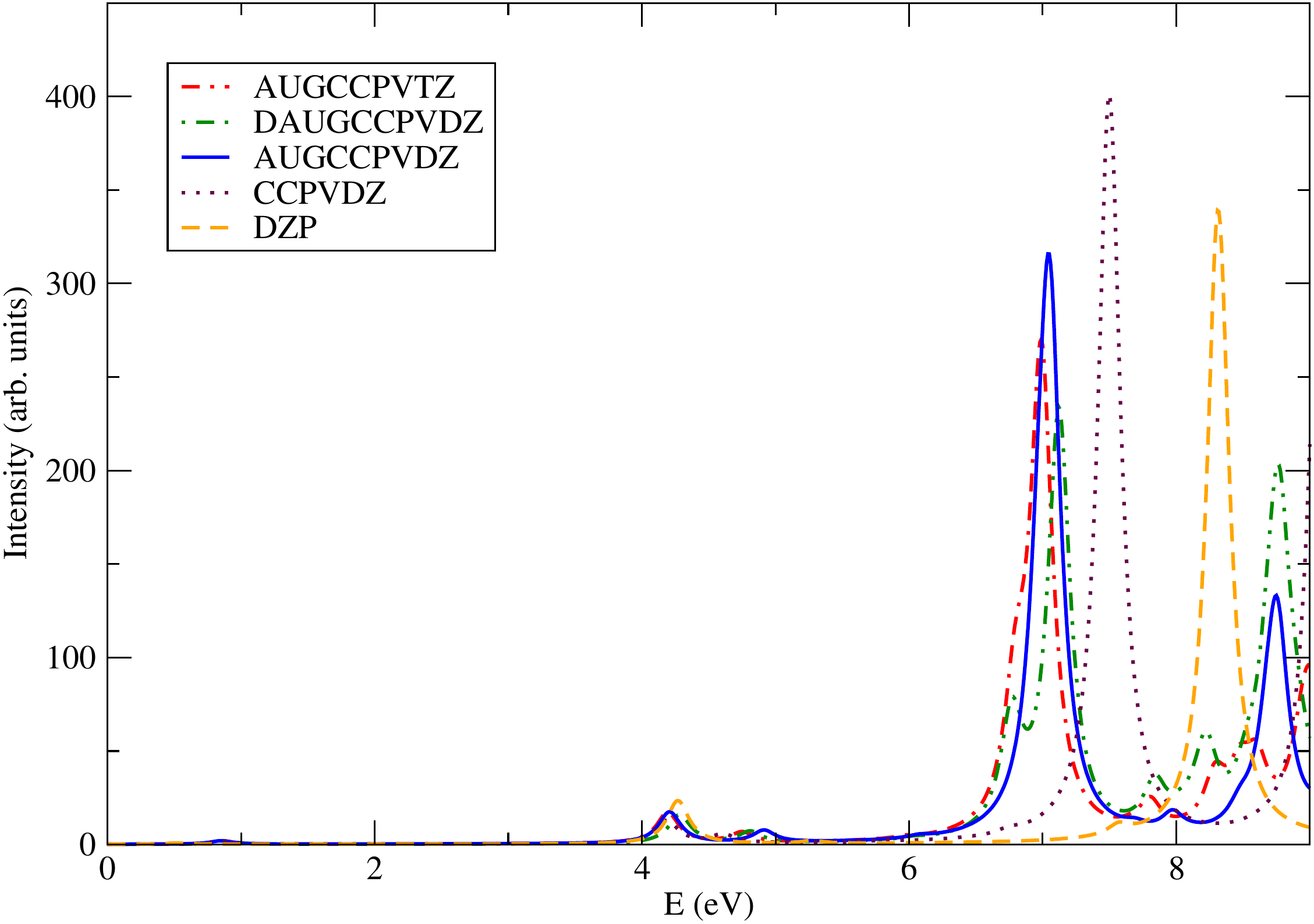,width=8.3cm,height=5.5cm}}
\caption{(Color online) Optical absorption in B$_{2}$ calculated using various
Gaussian contracted basis sets. Increasing more and more diffuse \textit{d}
type Gaussians shows negligible effect on optical spectra. }
\label{fig:basis} 
\end{figurehere}

\subsubsection{Orbital Truncation Schemes}

If the total number of orbitals used in a CI expansion is $N$, the
number of configurations in the calculation proliferates as $\approx N^{6}$,
which can become intractable for large values of $N$. Therefore,
it is very important to reduce the number of orbitals used in the
CI calculations. The occupied orbitals are reduced by employing the
so-called ``frozen-core approximation'' described earlier, while
the unoccupied (virtual) set is reduced by removing very high-energy
orbitals.

\begin{figurehere}
\centerline{\psfig{file=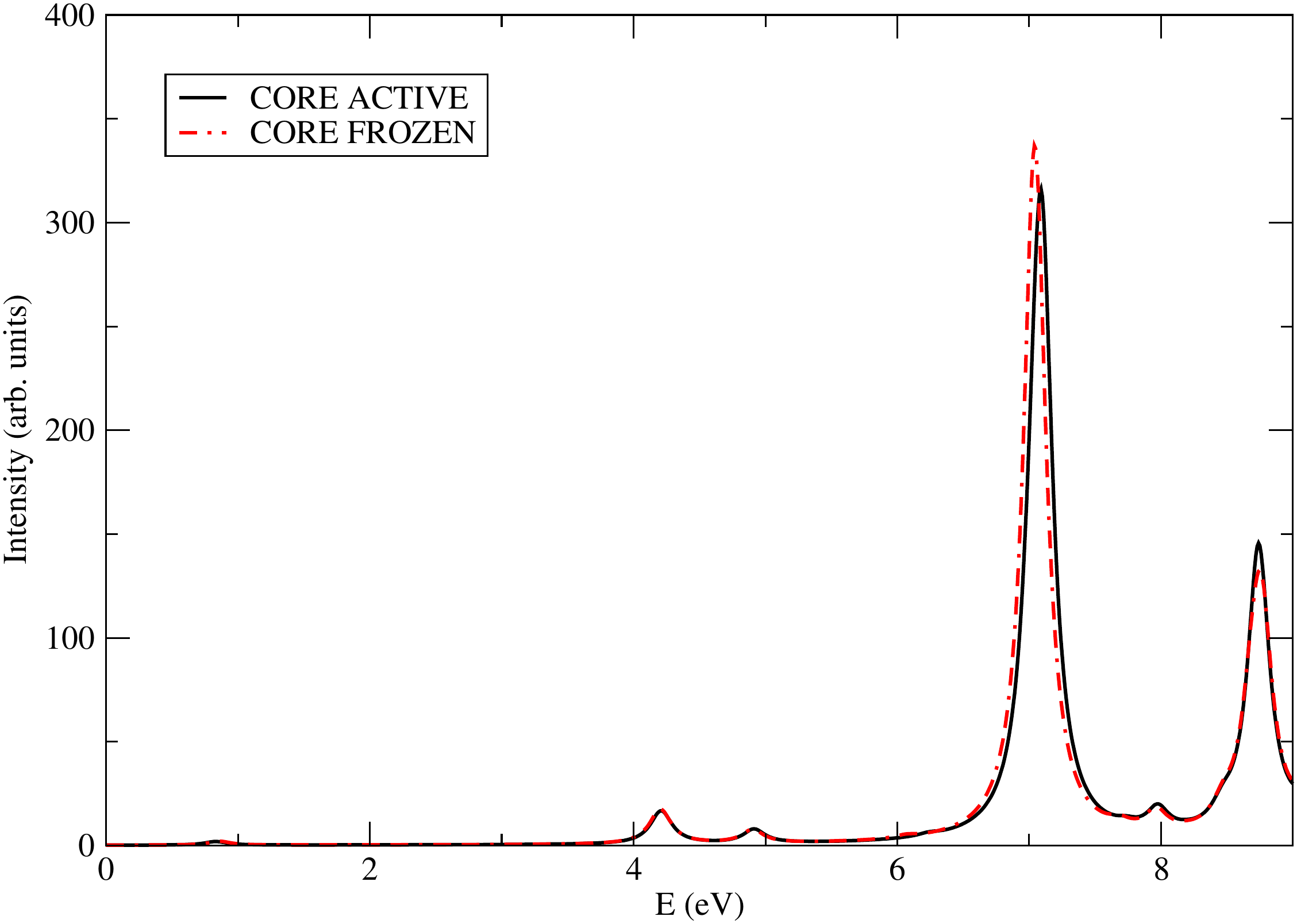,width=8.3cm,height=5.5cm}}
\caption{(Color online) The effect of freezing the core orbitals ($1s$) of
boron atoms on optical absorption spectrum of B$_{2}$. It renders
almost no effect on optical absorption spectrum.}
\label{fig:core} 
\end{figurehere}

The influence of freezing the $1s$ core orbitals on the optical absorption
spectrum of B$_{2}$ cluster is displayed in Fig. \ref{fig:core},
from which it is obvious that it makes virtually no difference to
the results whether or not the core orbitals are frozen. The effect
of removing the high-energy virtual orbitals on the absorption spectrum
of B$_{2}$ is examined in Fig. \ref{fig:nref}. From the figure it
is obvious that if all the orbitals above the energy of 1 Hartree
are removed, the absorption spectrum stays unaffected. Therefore,
in rest of the calculations, wherever needed, orbitals above this
energy cutoff were removed from the list of active orbitals. Theoretically
speaking this cutoff is sound, because we are looking for absorption
features in the energy range much smaller than 1 Hartree.

\begin{figurehere}
\centerline{\psfig{file=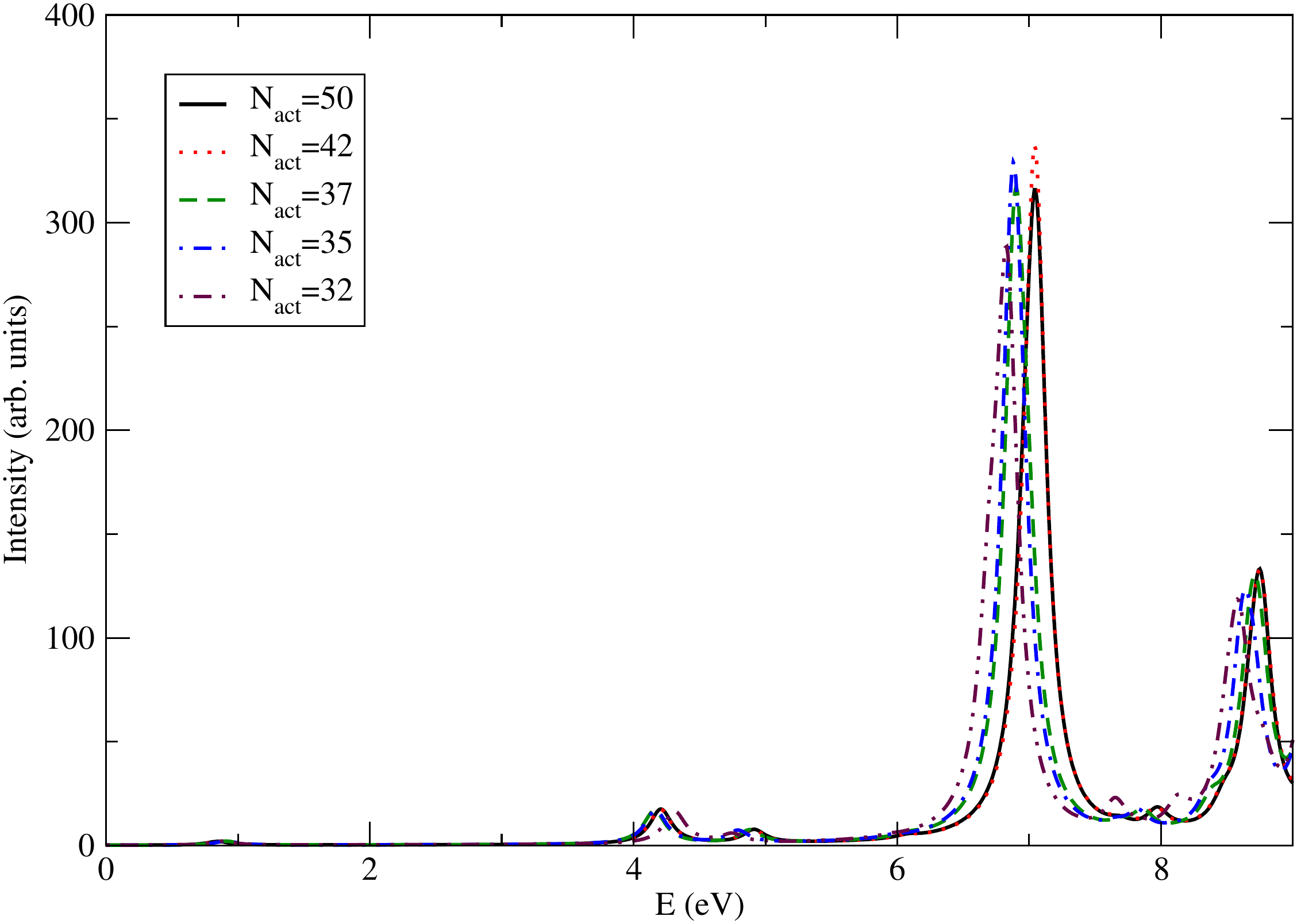,width=8cm,height=5.5cm}}
\caption{(Color online) The effect of the number of active orbitals (N$_{act}$)
on the optical absorption spectrum of B$_{2}$. Until N$_{act}$=42,
the optical spectrum does not exhibit any significant change. It corresponds
to 1.0 Hartree ($\approx27.2$ eV) virtual orbital energy. }
\label{fig:nref} 
\end{figurehere}

\subsubsection{Size of the CI expansion}

As mentioned earlier that the electron correlation effects in both
the ground and the excited states were accounted in our calculations
by including the relevant configurations in the reference list of
the MRSDCI expansion. The greater numerical accuracy demands the inclusion
of a large number of configurations in the reference list, but that
leads to a rapid growth in the size of the CI expansion, making the
calculations numerically prohibitive. However, here we are interested
in computing the energy differences rather than the absolute energies
of various states, for which good accuracy can be achieved even with
moderately large CI expansions. In Table \ref{tab:irrep} we present
the average number of reference states (N$_{ref}$) included in the
MRSDCI expansion and average number of configurations (N$_{total}$)
for different isomers. For a given isomer, the average has been calculated
across different irreducible representations which were needed in
these symmetry adapted calculations in order to compute the ground
and various excited states. The extensiveness of our calculations
can be seen from the number N$_{total}$, which is $\approx$ 77000
for the simplest cluster, and around four million for each symmetry
subspace of B$_{5}$. This makes us believe that our results are fairly
accurate.

Before we discuss the absorption spectrum for each isomer, we present
the ground state energies along with the relative energies of each
isomer are given in Table \ref{tab:energies}. The MRSDCI energy convergence
threshold was 10$^{-5}$ for all the isomers, with 10$^{-4}$ as convegrence
threshold for configuration coefficients. From the results it is obvious
that as far as the energetics are concerned, for the B$_{3}$ the
triangular structure is most stable, while for B$_{4}$ and B$_{5}$
the rhombus and pentagonal structures, respectively, are favorable.

\begin{tablehere}
\tbl{The average number of reference configurations (N$_{ref}$), and average
number of total configurations (N$_{total}$) involved in MRSDCI calculations
of various isomers of boron clusters.\label{tab:irrep}}
{\begin{tabular}{@{}cccc@{}}
 \toprule
 Cluster  & Isomer  & N$_{ref}$  & N$_{total}$ \\
 \colrule
 B$_{2}$  & Linear & 24 & 77245 \\
  &  &  & \\ 
 B$_{3}$  & Triangular  & 36 & 596798 \\
  & Linear & 41 & 671334 \\
  &  &  & \\
 B$_{4}$  & Rhombus  & 37 & 1127918 \\
  & Square  & 40 & 1070380 \\
  & Linear & 34 & 1232803 \\
  & Distorted Tetrahedron & 28 & 1253346 \\
  &  &  & \\
 B$_{5}$  & Pentagon  & 22 & 3936612 \\
  & Distorted Tri. bipyramid$^a$   & 7 & 3927508 \\ 
 \botrule
 \end{tabular}}  
 \begin{tabnote}
 $^a$ {C$_{s}$ symmetry of isomer converted to C$_{1}$ in calculations.} 
 \end{tabnote}
 \end{tablehere}

 \begin{tablehere}
 \tbl{Ground state (GS) energies (in Hartree) at MRSDCI level and the relative
 energies (in eV) of different isomers of clusters.\label{tab:energies}}
 {\begin{tabular}{@{}ccr@{\extracolsep{0pt}.}lc@{}}
 \toprule
 Cluster  & Isomer  & \multicolumn{2}{c}{GS energy} & \multicolumn{1}{c}{Relative}\tabularnewline
  &  & \multicolumn{2}{c}{(Ha)} & \multicolumn{1}{c}{energy (eV)}\tabularnewline
 \colrule
 B$_{2}$  & Linear  & -49&27844 & 0.00 \tabularnewline
  &  & \multicolumn{2}{c}{} & \tabularnewline
 B$_{3}$  & Triangular  & -73&98998 & 0.00 \tabularnewline
  & Linear & -73&92906 & 1.66\tabularnewline
  &  & \multicolumn{2}{c}{} & \tabularnewline
 B$_{4}$  & Rhombus  & -98&74004 & 0.00\tabularnewline
  & Square  & -98&73785 & 0.06\tabularnewline
  & Linear & -98&66575 & 2.02\tabularnewline
  & Distorted Tetrahedron & -98&63213 & 2.94\tabularnewline
  &  & \multicolumn{2}{c}{} & \tabularnewline
 B$_{5}$  & Pentagon  & -123&42652 & 0.00\tabularnewline
  & Distorted Tri. Bipyramid  & -123&31485 & 3.04\tabularnewline
 \botrule
 \end{tabular}}
 \end{tablehere}

\subsection{Calculated Photoabsorption Spectra of Various Clusters}

Next we present and discuss the results of our photoabsorption calculations
for each isomer.

\subsubsection{B$_{2}$}

The simplest and most widely studied cluster of boron is B$_{2}$
with D$_{\infty h}$ point group symmetry. Using the SDCI method,
we obtained its optimized bond length to be 1.59 \AA{} (\emph{cf.}
Fig. \ref{fig:geometries}\subref{fig:b2geom}), which is in excellent
agreement with the experimental value 1.589 \AA{}.\cite{herzberg_book}.
Using a DFT based methodology, Ati\c{s} \emph{et al.}\cite{turkish_boron},
obtained a bond length of 1.571 \AA{}, while Howard and Ray calculated
it to be 1.61 \AA{}, using the fourth-order perturbation theory (MP4).\cite{howard_ray}

Because the ground state of B$_{2}$ is a spin triplet, its many-particle
wave function predominantly consists of a configuration with two degenerate
singly-occupied molecular orbitals (SOMO) referred to as $H_{1}$
and $H_{2}$ in rest of the discussion. The excited state wave functions
will naturally consist of configurations involving electronic excitations
from the occupied MOs to the unoccupied MOs starting from lowest unoccupied
molecular orbital (LUMOs, $L$ for short). Our calculated photoabsorption
spectrum shown in Fig. \ref{fig:b2_lin_final} is characterized by
weaker absorptions at low energies, and a very intense one at high
energy. The many-particle wave functions of excited states contributing
to various peaks are presented in Table \ref{Tab:table_b2_lin}. A
feeble peak appears near 0.85 eV, dominated by $H_{2}\rightarrow L$
and $H_{2}\rightarrow L+4$ excitations compared to the HF reference
configuration. It is followed by a couple of smaller peaks at 4.20
and 4.91 eV. The most intense peak is found at 7.05 eV, to which two
closely spaced states contribute. Transition to the state near 6.97
eV is polarized transverse to the bond length, while the one close
to 7.05 eV carries the bulk of oscillator strength, and is reached
by longitudinally polarized photons. All these states exhibit strong
mixing of singly-excited configurations. Near 8 eV, a smaller peak
appears which has strong contributions from doubly-excited configurations
$H-1\rightarrow L$; $H_{1}\rightarrow L+2$ and $H-1\rightarrow L$;
$H_{2}\rightarrow L+2$. The wave functions of the excited states
contributing to all the peaks exhibit strong configuration mixing,
instead of being dominated by single configurations, pointing to the
plasmonic nature of the optical excitations.\cite{plasmon} 

\begin{figurehere}
\centerline{\psfig{file=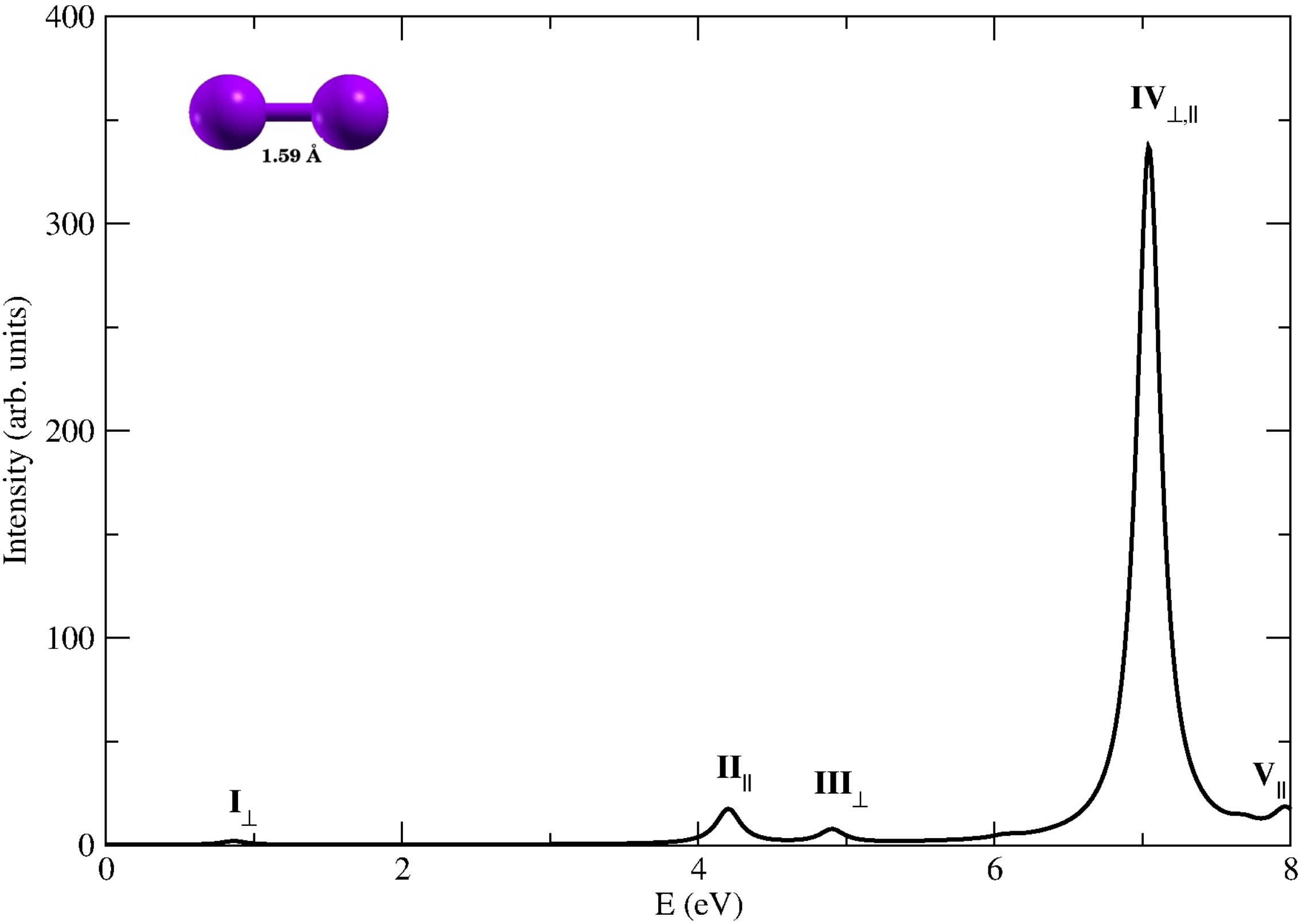,width=8.3cm}}
\caption{The linear optical absorption spectrum of B$_{2}$, calculated using
the MRSDCI approach. The peaks corresponding to the light polarized
along the molecular axis are labeled with the subscript $\parallel$,
while those polarized perpendicular to it are denoted by the subscript
$\perp$. For plotting the spectrum, a uniform linewidth of 0.1 eV
was used.}
\label{fig:b2_lin_final} 
\end{figurehere}

\subsubsection{B$_{3}$}

Boron trimer has two possible isomers, triangular and the linear one
shown in Figs. \ref{fig:geometries}\subref{fig:b3trgeom} and \ref{fig:geometries}\subref{fig:b3lingeom}.
We found equilateral triangle with D$_{3h}$ symmetry to be the most
stable isomer. The optimized bond length for triangular isomer is
1.55 \AA{}, with the ground state ($^{2}A_{1}^{'}$) energy 1.66 eV
lower than that of its linear counterpart. We also explored the possibility
of isosceles triangular structure as a favorable one, because B$_{3}$
is an open-shell system, making it a possible candidate for Jann-Teller
distortion. However, the CCSD optimised geometry corresponding to
the isosceles structure is so slightly different compared to the equilateral
one, that it is unlikely to affect the optical absorption spectrum
in a significant manner. Our calculated bond length is in good agreement
with experimental value 1.57 \AA{}\cite{hanley_whitten}, as well
as with other reported theoretical values of 1.553 \AA{},\cite{boustani_prb97}
1.56 \AA{} \cite{howard_ray} and 1.548 \AA{}.\cite{turkish_boron} 

The linear B$_{3}$ isomer with the D$_{\infty h}$ symmetry, and
the $^{2}\Sigma_{g}^{-}$ as ground state, was found to have equal
bond lengths. Our SDCI optimized bond length of 1.51 \AA{} agrees
well with the value 1.518 \AA{} reported by Ati\c{s} \emph{et al}.\cite{turkish_boron}

The photoabsorption spectra of two isomers of B$_{3}$ are presented
in Figs. \ref{fig:b3_lin_final} and \ref{fig:b3_tri_combined}. The
corresponding many-particle wave functions of excited states contributing
to various peaks are presented in Table \ref{Tab:table_b3_lin} and
\ref{Tab:table_b3_tri}. It is obvious that in the linear structure,
absorption begins at a lower energy as compared to the triangular
one, although the intensity of its low-energy peaks is very small.
In the triangular isomer on the other hand, most of the intensity
is concentrated at rather high energies, except for a weaker peak
close to 3 eV. The optical spectra of linear isomer begins with very
weak peaks at 0.7 eV (longitudinal polarization) and 2.7 eV (transverse
polarization), with their many-particle wave functions dominated by
singly-excited configurations. The relatively intense peak at 4.3
eV corresponding to a longitudinally polarized transition, is dominated
by doubly-excited configuraitons. It is followed by a small peak mainly
due to single excitation $H-2\rightarrow L$, near 5.9 eV. The most
intense peak of the spectrum occurs at 7.4 eV, followed by another
strong peak close to 7.7 eV. Both the features correspond to longitudinally
polarized transitions, with the many particle wave functions of the
concerned states being strong mixtures of single and double excitations
with respect to the HF reference state. We note that in the absorption
spectrum of the linear cluster, quite expectedly, the bulk of the
oscillator strength is carried by longitudinally polarized transitions.

Because the triangular cluster is a planar cluster, its orbitals can
be classified as in-plane $\sigma$ orbitals, and the out-of-plane
$\pi$ orbitals. Both the HOMO (a singly occupied orbital, in this
case) and the LUMO for this isomer are $\sigma$-type orbitals. For
this system, two types of optical absorptions are possible: (a) those
polarized in the plane of the cluster, and (b) the ones polarized
perpendicular to that plane. Our calculations reveal that the transitions
correponding to perpendicular polarization ($z$ direction), except
for a couple of peaks, have negligible intensities. From Fig. \ref{fig:b3_tri_combined}
it is obvious that the optical absorption in the triangular isomer
starts with a very weak $z-$polarized feature near 0.8 eV (peak I),
corresponding  to a state with the wave function dominated by single
excitations ($\pi\rightarrow\sigma^{*}$).  This is followed by a
series of peaks ranging from II to VI which correpsond to the photons
polarized in the plane of the cluster. All these peaks are dominated
by states consisting primarily of singly-excited configurations of
the $\sigma\rightarrow\sigma^{*}$ type. The most intense peak VI
is followed by a shoulder-like feature (VII) corresponding to a $z$-polarized
absorption. 

If we compare the absorption spectra of the linear and the triangular
B$_{3}$, the peak at 4.34 eV in the spectrum of the linear
cluster is the distinguishing feature, and can be used to differentiate
between the two isomers. 

\begin{figurehere}
\centerline{\psfig{file=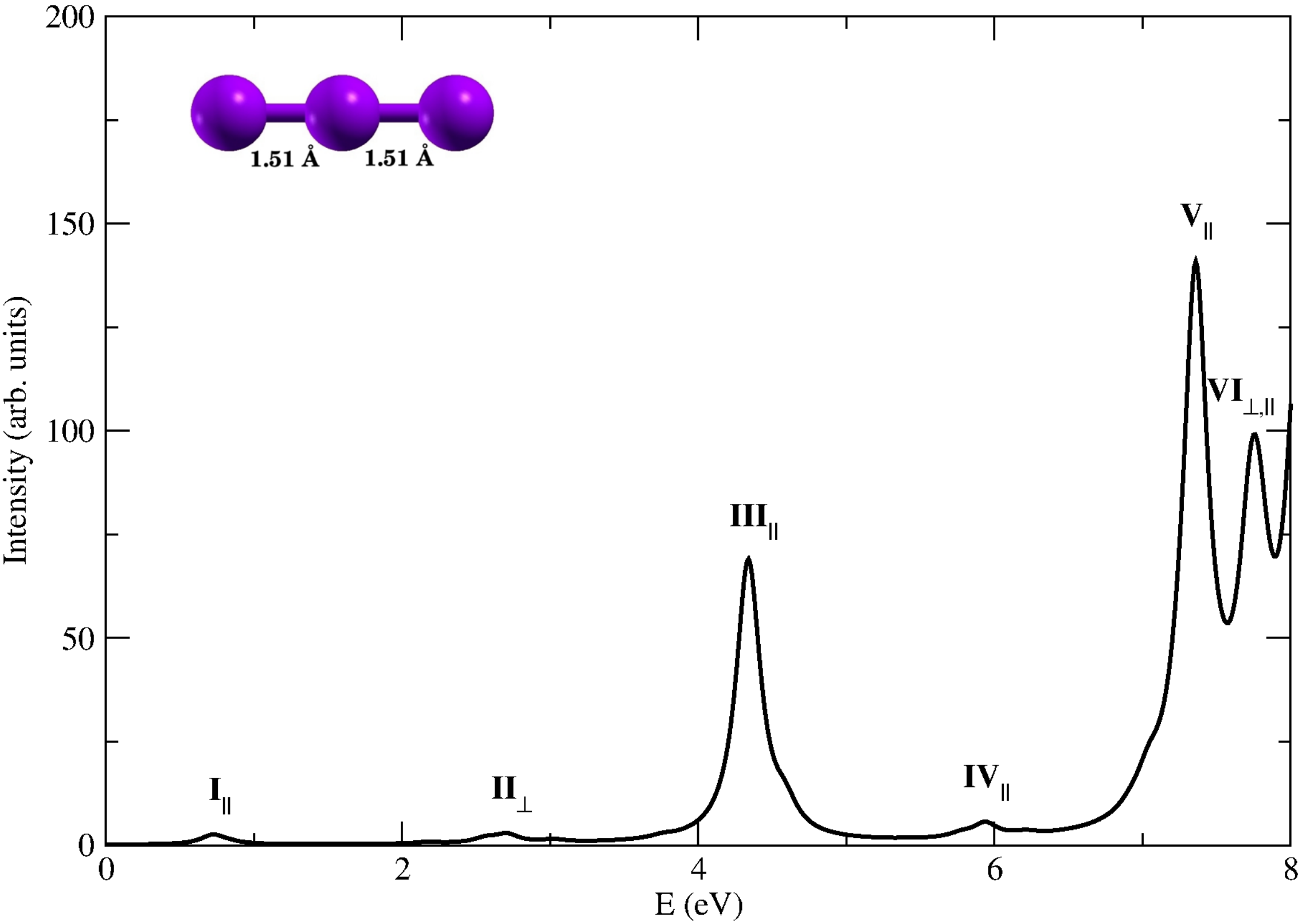,width=8.3cm}}
\caption{The linear optical absorption spectrum of linear B$_{3}$, calculated
using the MRSDCI approach. The peaks corresponding to the light polarized
along the molecular axis are labeled with subscript $\parallel$,
while those polarized perpendicular to it are denoted by the subscript
$\perp$. For plotting the spectrum, a uniform linewidth of 0.1 eV
was used.}
\label{fig:b3_lin_final} 
\end{figurehere}

\begin{figurehere}
\centerline{\psfig{file=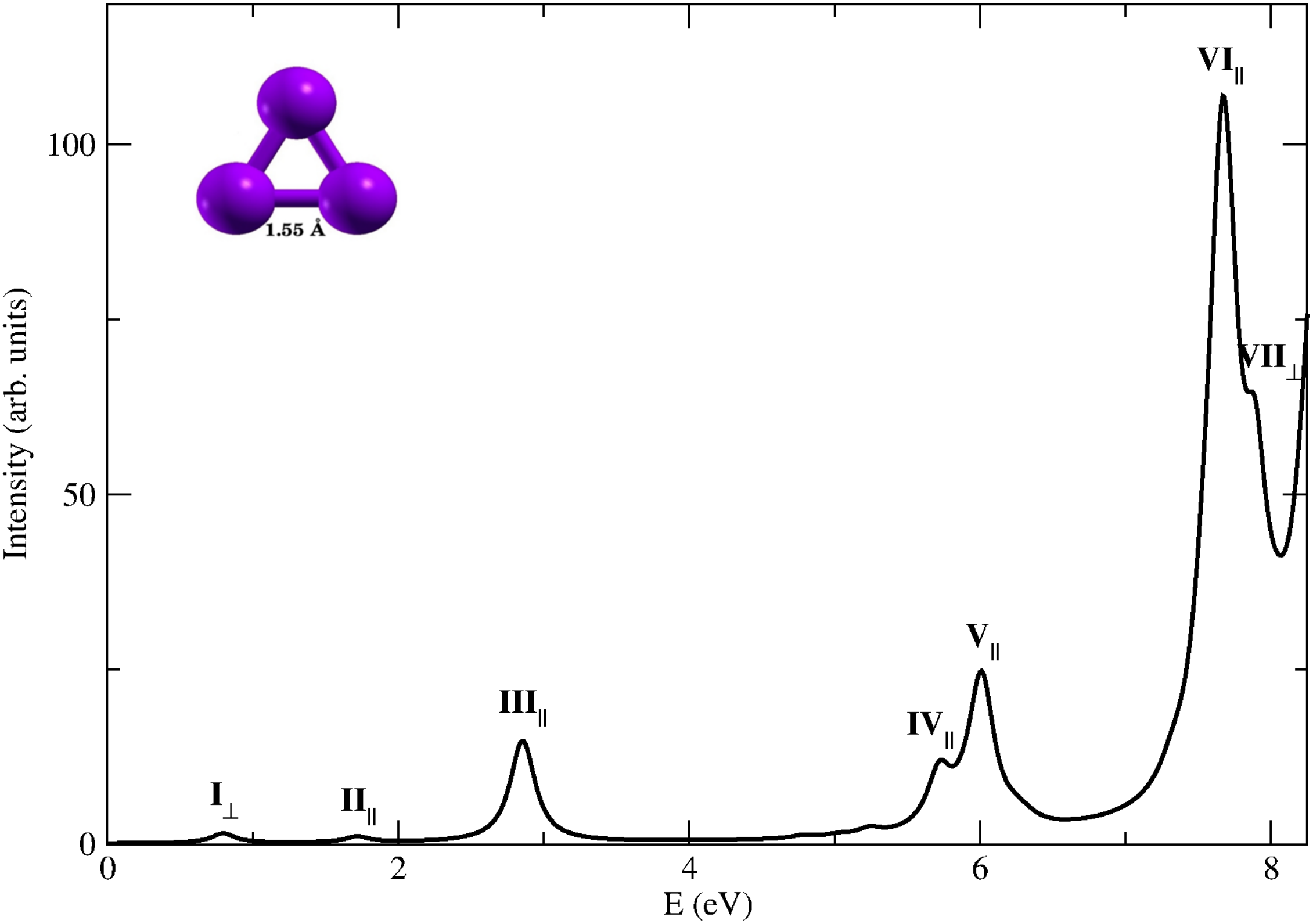,width=8.3cm}}
\caption{(Color online) The linear optical absorption spectrum of triangular
B$_{3}$ calculated using the MRSDCI approach. Peaks corresponding
to light polarized in the plane of the molecule are labeled with subscript
$\parallel$, while those polarized perpendicular to the plane are
denoted by the subscript $\perp$. For plotting the spectrum, a uniform
linewidth of 0.1 eV was used.}
\label{fig:b3_tri_combined} 
\end{figurehere}

\subsubsection{B$_{4}$}

For the B$_{4}$ cluster, we investigated the rhombus, square, linear
and tetrahedral structures. While the rhombus shaped isomer was found
to have the lowest energy, but the square isomer is higher in energy
only by a small amount. As a matter of fact, at the HF level the energies
of the two isomers were found to be almost degenerate. It was only
after the electron correlation effects were included at the CI level
that the rhombus stabilized by $\approx$0.06 eV (\emph{cf.} Table
\ref{tab:energies})with respect to the square. For the rhombus, the
ground state was $^{1}A_{g}$, with the optimized bond length 1.529
\AA{}, and the short diagonal length 1.909 \AA{}. These results are
in good agreement with with the correponding lengths of 1.528 \AA{}
and 1.885 \AA{} reported by Boustani,\cite{boustani_prb97}, and 1.523
\AA{} and 1.884 \AA{} computed by Ati\c{s} \emph{et al.}\cite{turkish_boron}
Both HOMO and LUMO of rhombus isomer are $\sigma$ orbitals.

For the square isomer, with D$_{4h}$ symmetry, the electronic ground
state is expectedly $^{1}$A$_{1g}$. As shown in Fig. \ref{fig:geometries}\subref{fig:b4sqrgeom},
our optimized bound length is 1.53 \AA{}, which agrees well with the
values 1.527 \AA{} and 1.518 \AA{} as reported in Refs. \refcite{boustani_prb97} and \refcite{turkish_boron}.
In this isomer, HOMO is a $\sigma$ orbital while LUMO is a $\pi$
orbital.

Linear B$_{4}$, with the D$_{\infty h}$ symmetry, has the electronic
ground state of $^{1}\Sigma_{g}^{-}$. However, energetically linear
structure is 2.02 eV higher than the rhombus one (\emph{cf.} Table
\ref{tab:energies}) which rules out its existence at the room temperatures.
As per Fig. \ref{fig:geometries}\subref{fig:b4lingeom}
, the central
bond length was found to be 1.49 \AA{}, with the two outer bonds being
1.55 \AA{} in length. For the same bonds, Ati\c{s} \emph{et al.} reported
these lengths to be 1.487 \AA{} and 1.568 \AA{}, respectively.\cite{turkish_boron}

The distorted tetrahedral structure having C$_{3v}$ symmetry, made
up of four isosceles triangular faces with lenghts 1.785 \AA{}, 1.785
\AA{} and 1.512 \AA{}. This isomer also lies much higher in energy
as compared to the most stable rhombus structure.

The absorption spectra of rhombus, square, linear, and tetrahedral
isomers are presented in Figs. \ref{fig:b4_rho_combined}, \ref{fig:b4_sqr_combined},
\ref{fig:b4_lin_final}, and \ref{fig:b4_tetra_combined} 
respectively.
From the figures it is obvious that the general features of the absorption
spectra of rhombus and square isomers are similar, except that the
rhombus spectrum, with the onset of the absorption near 4 eV, is red-shifted
by about 1 eV as compared to the square. The absorption spectrum of
the linear structure is slightly red-shifted as compared to the rhombus
and square shaped isomers, with the majority of absorption occuring
in the energy range 5--8 eV. This aspect of the photoabsorption in
B$_{4}$ is similar to the case of B$_{3}$ for which also the linear
structure exhibited a redshifted absorption compared to the triangular
one.

Since B$_{4}$ rhombus isomer has $D_{2h}$ symmetry, we can represent
the absorption due to light polarized in different directions in terms
of irreducible representations of $D_{2h}$. So absorption due to
in-plane polarized light corresponds to $B_{1u}$ and $B_{2u}$, while
$B_{3u}$ corresponds to light polarized in the direction perpendicular
to the plane of the isomer. 

The polarization resolved absorption spectrum of rhombus B$_{4}$,
as shown in Fig. \ref{fig:b4_rho_combined}, exhibits a rather blue-shifted
nature as compared to the linear isomer. The many-particle wave functions
of excited states contributing to various peaks are presented in Table
\ref{Tab:table_b4_rho}. The onset of spectrum is seen at 4.15 eV
followed by a peak at around 6.12 eV. Both of them are due to $y-$polarized
component, i.e. along the larger diagonal. The dominant contribution
to these peaks come from $\sigma\rightarrow\pi^{*}$ for former, and
$\pi\rightarrow\pi^{*}$ for latter. The $x$-component does not contribute
much in the whole spectrum, except for minor peaks at 4.2 eV and 6.6
eV. It is characterised by mainly $\pi\rightarrow\pi^{*}$ type transitions.
It is followed by a relatively low intensity peak at 7.3 eV due to
$y$-polarized component with leading contribution from $\sigma\rightarrow\pi^{*}$
transitions. The most intense peak, at 7.84 eV, having $y$-polarization
component, is characterized by $\sigma\rightarrow\pi^{*}$ type of
transitions. There are no direct $H\rightarrow L$ transitions for
this isomer, because they are dipole forbidden. The absorption due
to light polarized in the direction perpendicular to the plane of
isomer is negligible.

\begin{figurehere}
\centerline{\psfig{file=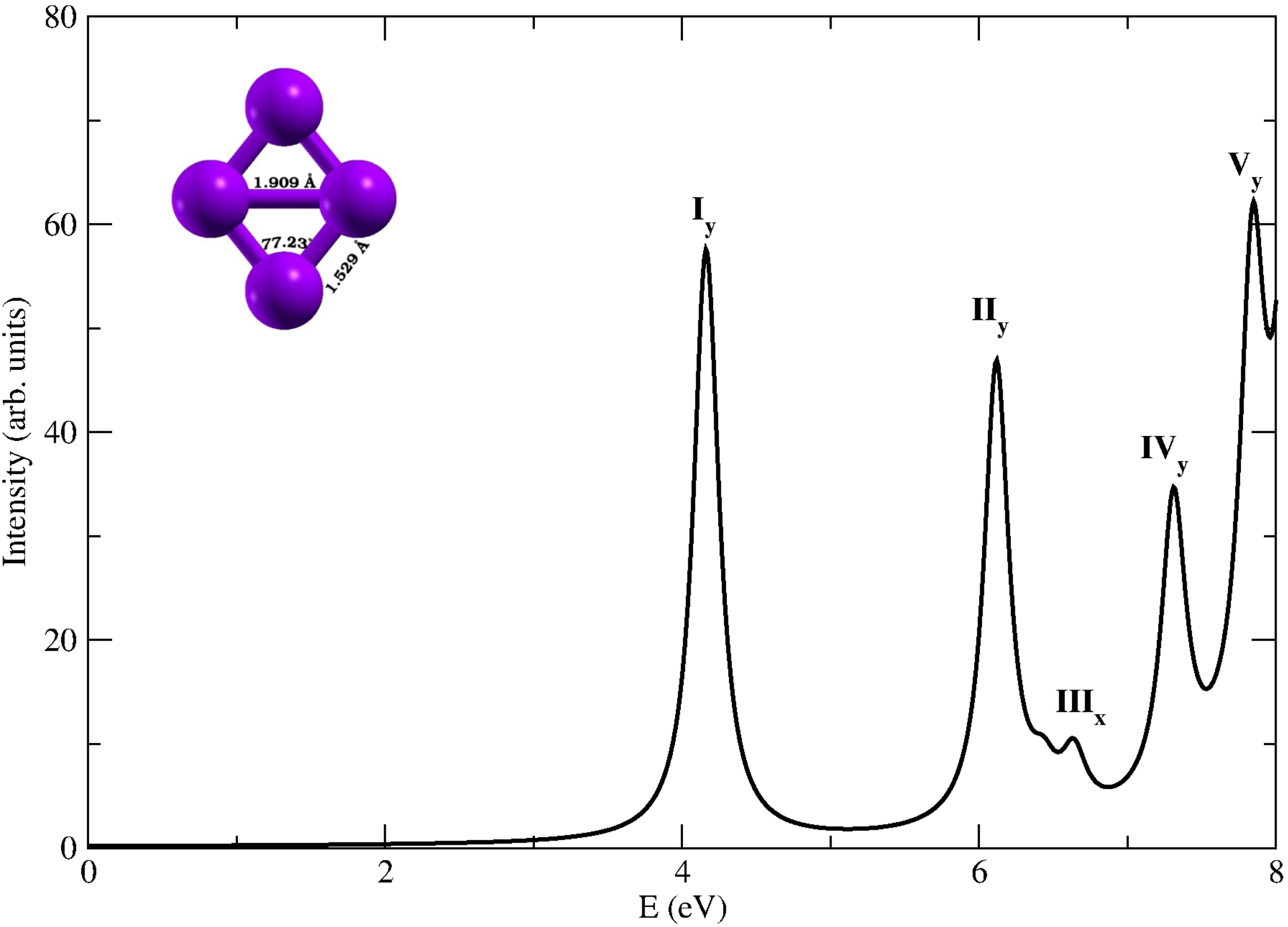,width=8.3cm}} 
\caption{(Color online) The linear optical absorption spectrum of B$_{4}$
rhombus geometry using the MRSDCI approach. Isomer is aligned in $x-y$
plane with short diagonal along $x$-axis. Peaks corresponding to
light polarized along $x$ and $y$-axis are labeled with subscript
$x$ and $y$. For plotting the spectrum, a uniform linewidth of 0.1
eV was used.}
\label{fig:b4_rho_combined} 
\end{figurehere}

The square B$_{4}$ isomer, because of its symmetry, gets equal contribution
to absorption spectrum from both $x-$ and $y-$component. It corresponds
to in-plane polarization due to $B_{1u}$ and $B_{2u}$ irreducible
representation, while $B_{3u}$ corresponds to light polarized in
the direction perpendicular to the plane of the isomer. However, in
this isomer also, the contribution due to latter is quite negligible.
The many-particle wave functions of excited states contributing to
various peaks are presented in Table \ref{Tab:table_b4_sqr}. It shows
just one major peak at 4.88 eV below 7 eV, characterized by $\sigma\rightarrow\pi^{*}$;$\sigma\rightarrow\pi^{*}$
double excitation. Two smaller peaks appear in this range at 5.5 eV
and 6.4 eV, with leading contributions from $\sigma\rightarrow\pi^{*}$;$\sigma\rightarrow\pi^{*}$
and $\sigma\rightarrow\pi^{*}$;$\pi\rightarrow\pi^{*}$ excitations
respectively. Beyond 7 eV, there are many closely spaced peaks including
the most intense one at 7.89 eV. It is characterized by double excitation
$\sigma\rightarrow\pi^{*}$;$\pi\rightarrow\pi^{*}$. In this isomer
also, a direct $H\rightarrow L$ transitioin is forbidden. Though,
there is very little difference in total energies of rhombus and square
isomers of B$_{4}$, their optical absorption spectra are completely
different. They can be easily identified from each other by looking
at number of peaks below 7 eV energy. Rhombus exhibits two major peaks,
while square has just one. 

\begin{figurehere}
\centerline{\psfig{file=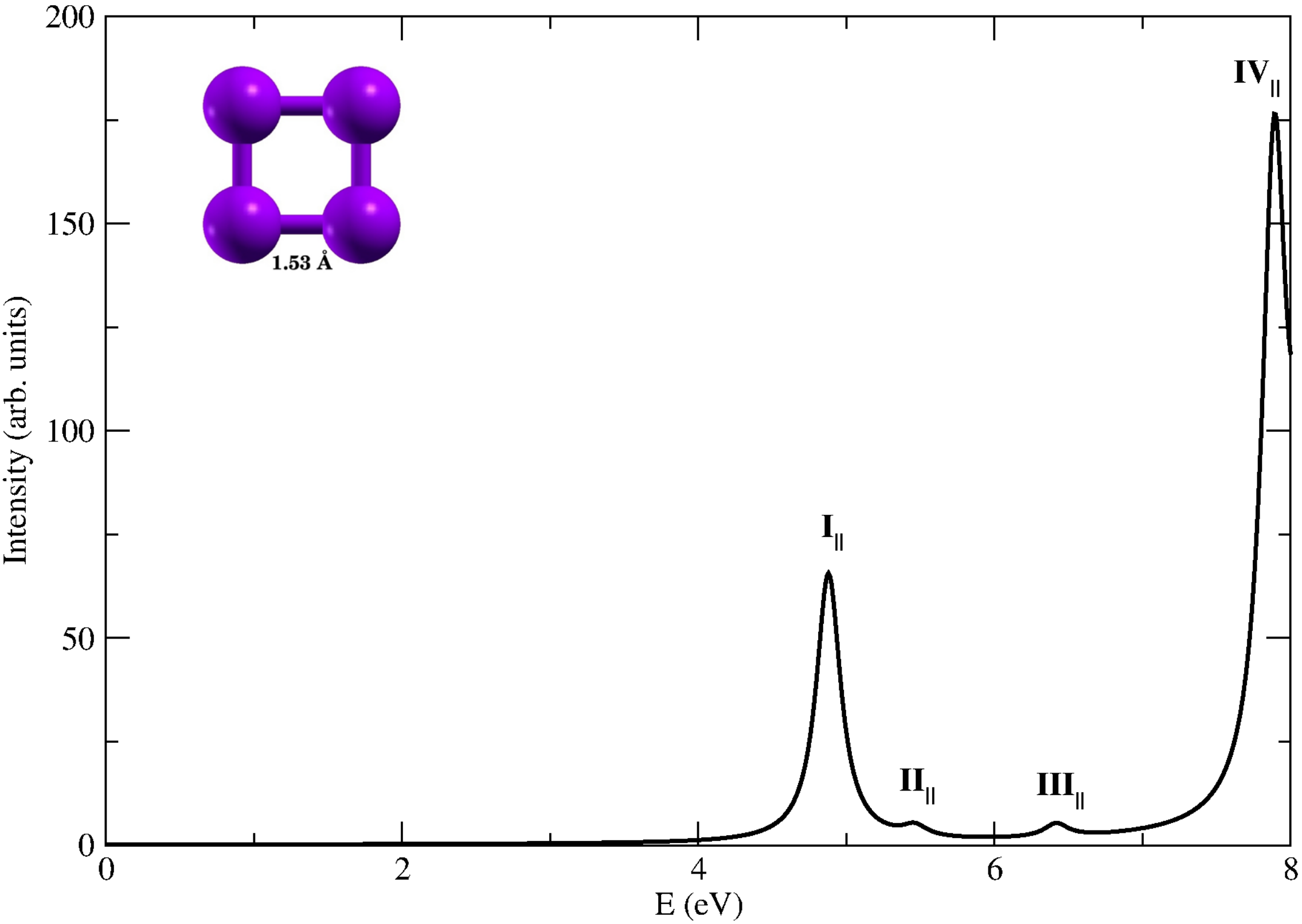,width=8.3cm}}
\caption{(Color online) The linear optical absorption spectrum of B$_{4}$
square geometry using MRSDCI approach. Isomer is aligned in $x-y$
plane. Spectrum represents the equal contribution from light polarized
in $x$ and $y$ direction. Peaks corresponding to light polarized
in the plane of the molecule are labeled with subscript $\parallel$.
For plotting the spectrum, a uniform linewidth of 0.1 eV was used.}
\label{fig:b4_sqr_combined} 
\end{figurehere}

Linear B$_{4}$ isomer exhibits absorption with few, but sharp peaks.
The many-particle wave functions of excited states contributing to
various peaks are presented in Table \ref{Tab:table_b4_lin}. The
onset of optical absorption occurs near 4.5 eV, due to absorption
of long-axis polarized light, followed by two major peaks at 5.95
eV, and 7.36 eV. The first of these two intense peaks, peak II is
dominated by singly-excited configurations, while the second one (peak
III) is a strong mixture of both singly- and doubly-excited configurations
with respect to the HF reference configuration. 

\begin{figurehere}
\centerline{\psfig{file=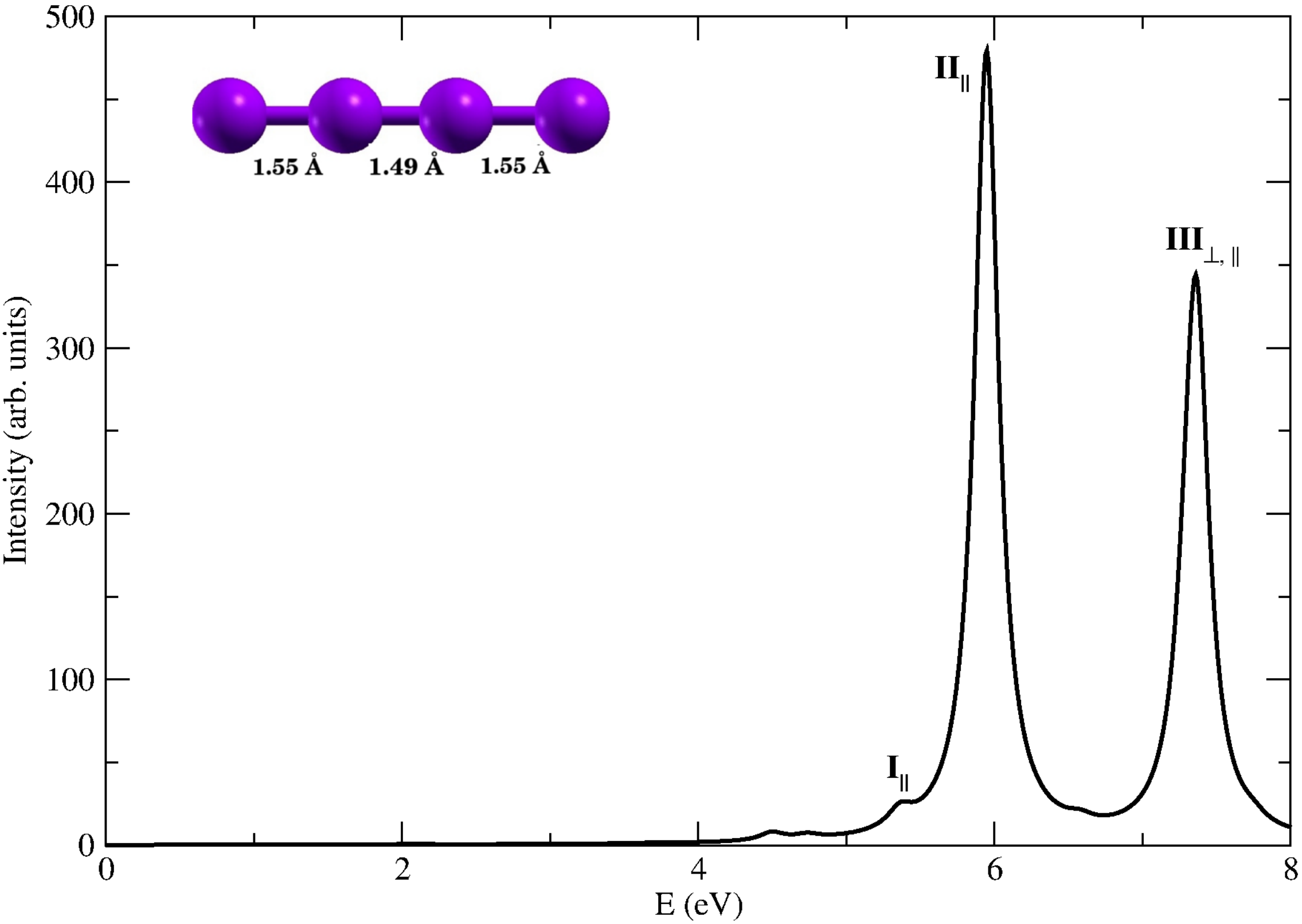,width=8.3cm}}
\caption{The optical absorption spectrum of linear B$_{4}$, calculated using
the MRSDCI approach. The peaks corresponding to the light polarized
along the molecular axis are labeled with subscript $\parallel$,
while those polarized perpendicular to it are denoted by the subscript
$\perp$. For plotting the spectrum, a uniform linewidth of 0.1 eV
was used.}
\label{fig:b4_lin_final} 
\end{figurehere}

The 3D structure, a distorted tetrahedral isomer, exhibits an absorption
spectrum very different from other isomers, as displayed in Fig. \ref{fig:b4_tetra_combined}.
The many-particle wave functions of excited states contributing to
various peaks are presented in Table \ref{Tab:table_b4_tetra}. It
is the only B$_{4}$ isomer to exhibit peaks below 4 eV. The absorption
spectrum is spread over a much larger energy range, and is almost
continuous. The oscillator strengths associated with various peaks
are much smaller than in other isomers, and most of the peaks appear
pairwise. The onset of absorption spectrum is seen at around 1.1 eV,
characterised mainly by an excited dominated by single-excitation
$H\rightarrow L$ (\emph{cf.} Table \ref{Tab:table_b4_tetra}). In
this isomer, in contrast to other B$_{4}$ isomers, direct $H\rightarrow L$
transitions are allowed. Higher energy peaks in this isomer are dominated
by doubly-excited configurations, and, are, therefore, sensitive to
the electron-correlation effects.

\begin{figurehere}
\centerline{\psfig{file=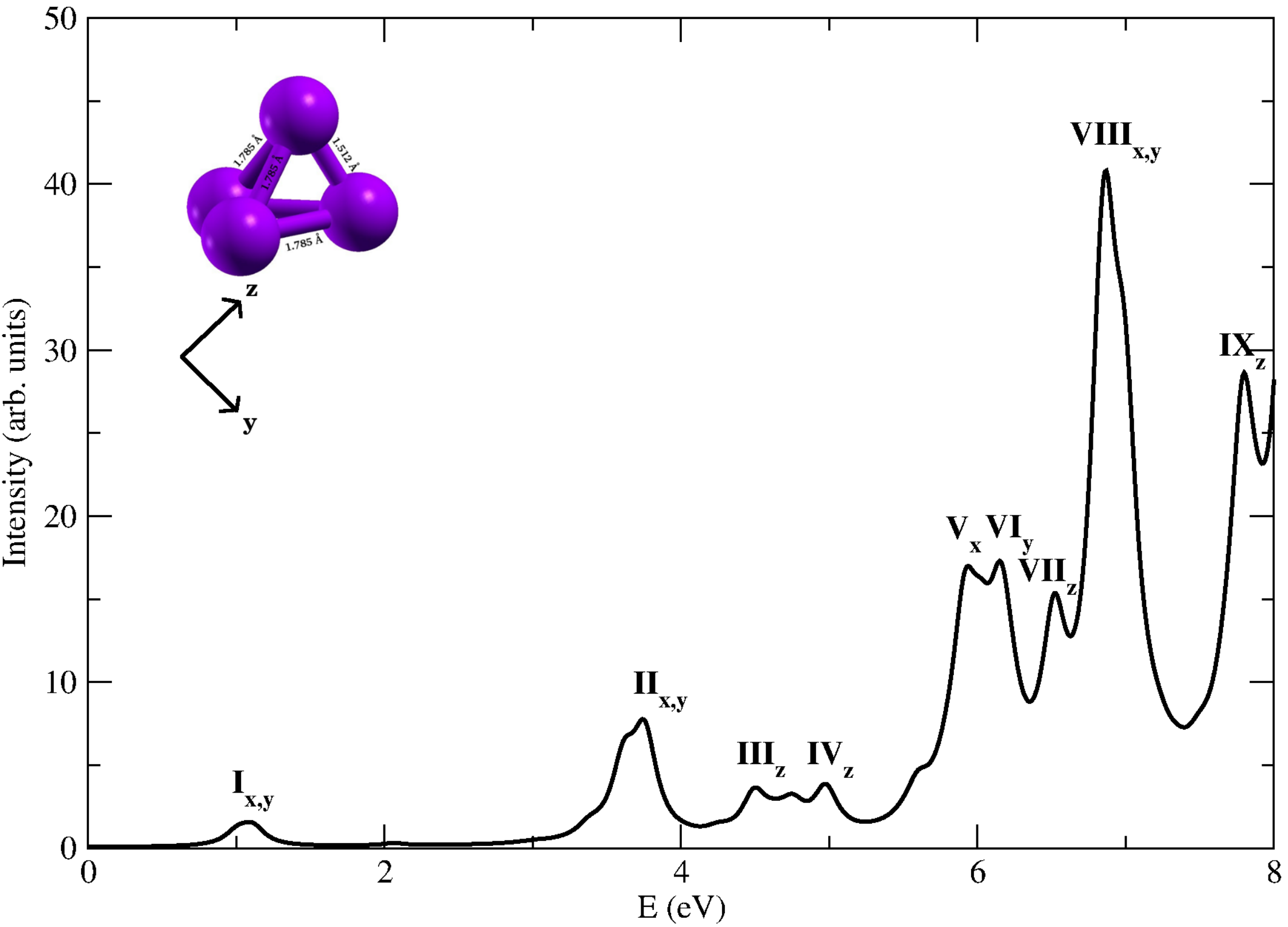,width=8.3cm}}
\caption{(Color online) The linear optical absorption spectrum of B$_{4}$
distorted tetrahedral geometry using the MRSDCI approach. Peaks corresponding
to light polarized along $x$, $y$ and $z$-axis are labeled with
subscript $x$, $y$ and $z$ respectively. For plotting the spectrum,
a uniform linewidth of 0.1 eV was used.}
\label{fig:b4_tetra_combined} 
\end{figurehere}

\subsubsection{B$_{5}$}

We investigated two isomers of B$_{5}$: a Jahn-Teller distorted pentagon
with the C$_{2v}$ symmetry, and (b) a triangular bipyramid with the
C$_{s}$ point group symmetry. The latter one is the second 3-D structure
of the boron clusters probed in this work. The lowest lying pentagon
isomer, has $^{2}$B$_{2}$ electronic ground state, and is 3.04 eV
lower in energy as compared to the bipyramid structure. For the pentagon,
the symmetry of ground state at the SCF level was A$_{1}$, however,
at the MRSDCI level the B$_{2}$ state became lower in energy, in
agreement with the previous calculations of Boustani.\cite{boustani_prb97}
Our optimized geometry for the pentagon (Fig. \ref{fig:geometries}\subref{fig:b5pengeom})
corresponds to an average bond length of 1.56 \AA{}, as against 1.57
\AA{} reported by Boustani\cite{boustani_prb97}, and 1.644 \AA{}
reported by Ati\c{s} \emph{et al}.\cite{turkish_boron}. The singly
occupied molecular orbital (denoted by $H$) and LUMO of pentagon
isomers are of $\pi$ and $\sigma$ type, respectiely. The bond lengths
for the bipyramid structure are shown in Fig. \ref{fig:geometries}\subref{fig:b5tetgeom},
with an average bond length of 1.704 \AA{}. The triangular base was
found to be isosceles with 1.97 \AA{} as equal sides, and 1.75 \AA{}
as the other side. 

The absorption spectra of the two isomers are presented in Figs. \ref{fig:b5_pen_combined}
and \ref{fig:b5_tet_combined}. The many-particle wave functions of
excited states contributing to various peaks are presented in Table
\ref{Tab:table_b5_pen} and \ref{Tab:table_b5_tet} respectively.
From the figures it is obvious that the absorption in the bipyramid
starts at much lower energies as compared to the pentagonal isomer.
Intense absorption peaks in pentagon B$_{5}$ are located at energies
higher than 5 eV, with three equally intense peaks at 5.58 eV, 6.30
eV and 7.16 eV, with the photons polarized along the plane of the
molecule direction. It has an underlying low intensity absorption
contribution from photons polarized along $z-$ direction, which is
perpendicular to the molecular plane. The major contribution to the
peak at 5.58 eV comes from $\pi\rightarrow\sigma^{*}$ and $\sigma\rightarrow\sigma^{*}$,
single excitations. The latter configuration also contributes to the
most intense peak at 6.30 eV. The peak at 7.16 eV is mainly due to
$\sigma\rightarrow\sigma^{*}$ type transitions.

\begin{figurehere}
\centerline{\psfig{file=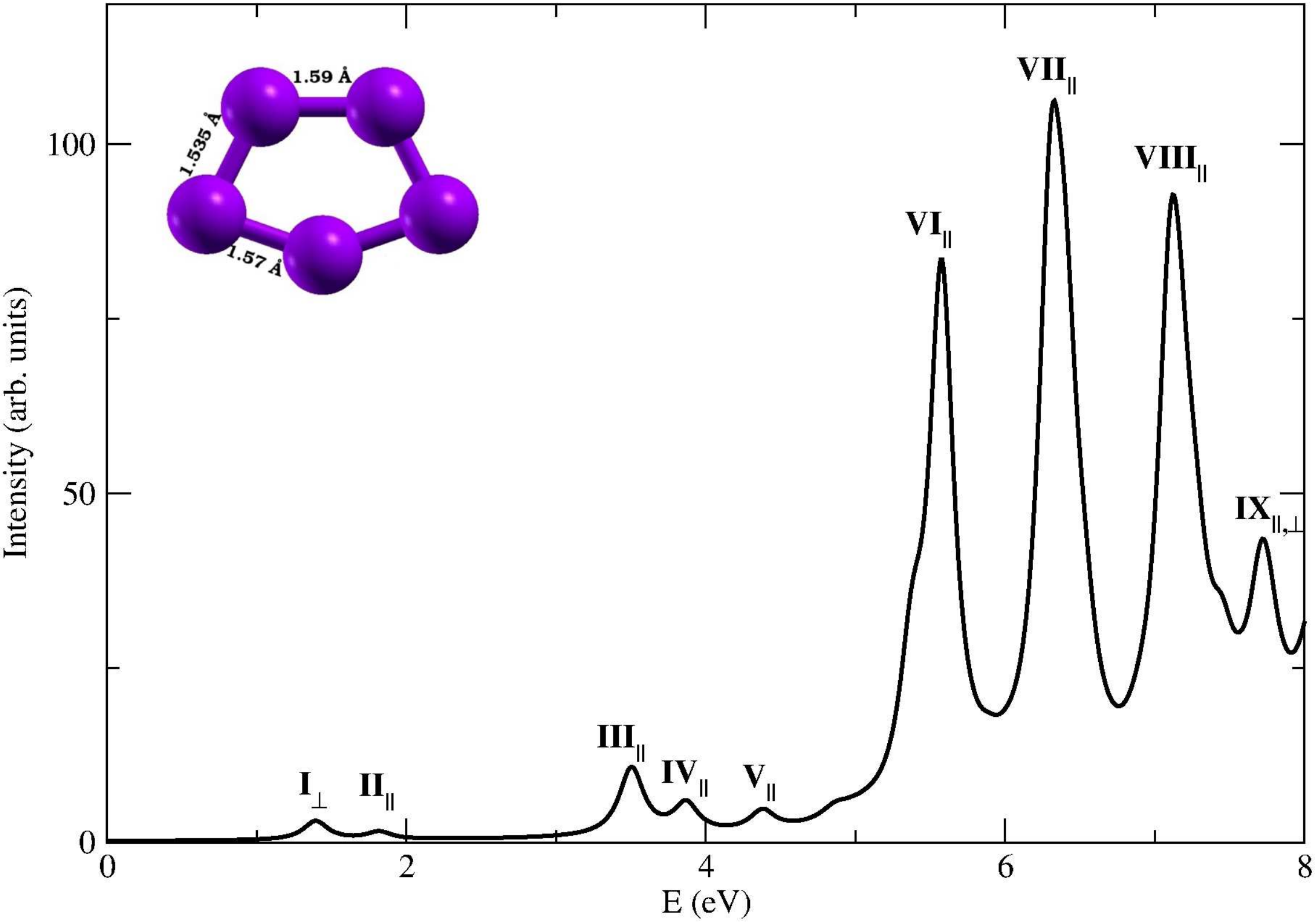,width=8.3cm}}
\caption{(Color online) The linear photo-absorption spectrum of pentagon B$_{5}$,
calculated using the MRSDCI approach. The peaks corresponding to the
light polarized in the plane of the molecule are labeled with subscript
$\parallel$, while those polarized perpendicular to it are denoted
by the subscript $\perp$. For plotting the spectrum, a uniform linewidth
of 0.1 eV was used.}
\label{fig:b5_pen_combined} 
\end{figurehere}

Since the B$_{5}$ triagonal bipyramid isomer is not a symmetric one,
the calculations were done using C$_{1}$ symmetry, thereby increasing
the difficulty in diagonalizing the Hamiltonian. Hence, in order to
reduce the matrix size, we have used a smaller number of reference
configurations, and also relaxed the energy convergence threshold
criterion a little.

The optical absorption spectrum of B$_{5}$ triangular bipyramid isomer
is exhibited by almost equally spaced peaks at relatively lower energies.
The optical absorption starts at 1.74 eV characterized by $H-1\rightarrow L+4$
configuration. It is followed by two equal intensity peaks at 3.16
eV and 4.27 eV with contributions from single excitations $H-2\rightarrow H\mbox{ and }H-1\rightarrow L+2$,
respectively. The most intense peak is found at 7.52 eV dominated
by the doubly excited configuration $H-2\rightarrow H\:;\: H-1\rightarrow L+2$.
There are two distinguishing features as far as the optical absorption
in the two isomers is concerned: (a) presence of three intense peaks
in the higher energy region of the absorption spectrum of the pentagonal
isomer, and (b) occurance of equally spaced absorption peaks at lower
energies in the spectrum of bipyramidal isomer.

\begin{figurehere}
\centerline{\psfig{file=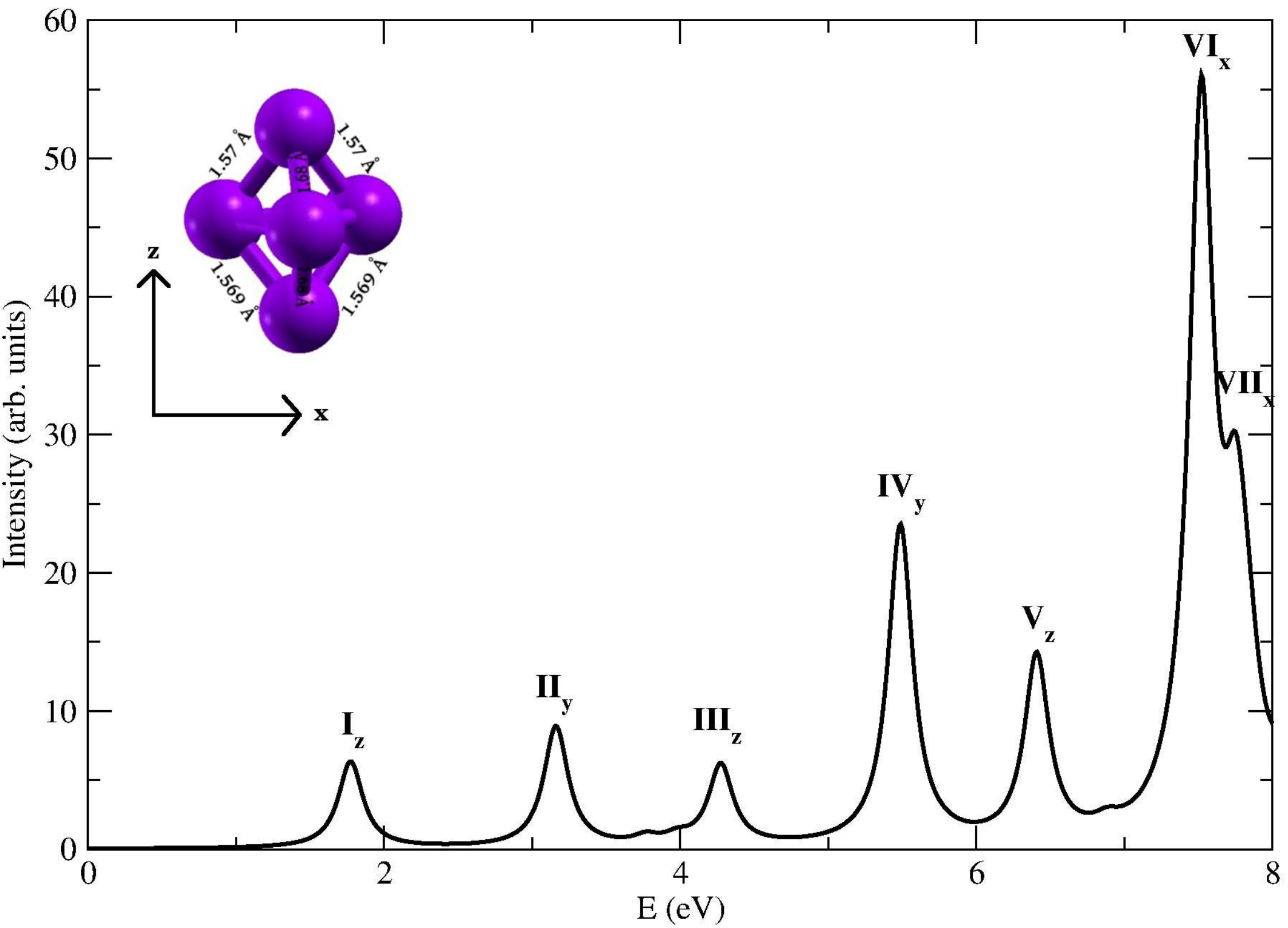,width=8.3cm}}
\caption{(Color online) The linear photo-absorption spectrum of distorted triangular
bipyramid B$_{5}$, calculated using the MRSDCI approach. Peaks corresponding
to light polarized along $x$, $y$ and $z$-axis are labeled with
subscript $x$, $y$ and $z$ respectively. For plotting the spectrum,
a uniform linewidth of 0.1 eV was used.}
\label{fig:b5_tet_combined} 
\end{figurehere}

\section{Conclusions and Outlook}

\label{sec:conclusions}

We presented systematic large-scale all-electron correlated calculations
of photoabsorption spectra of boron clusters B$_{n}$, ($n=$2--5)
with several possible isomers of each cluster. The calculations were
perfomred using the MRSDCI method which takes electron correlations
into account at a sophisticated level, both for the ground and the
excited states. For a cluster consisting of a given number of atoms,
significant changes were observed in absorption spectra for different
isomers, indicating a strong structure-property relationship. Therefore,
our computed spectra can be used in the future photoabsorption experiments
to distinguish between different isomers of a cluster, something which
is not possible with the conventional mass spectrometry. We also analyzed
the many-particle wave functions of various excited states and found
them to be a mixture of a large number of configurations, indicating
the nature of photoexcited states in these clusters to be plasmonic.\cite{plasmon}
A noteworthy aspect of the groundstate photoabsorption of various
clusters was the absence of high-intensity peaks in the low-energy
region of the spectrum. The most intense peaks occurred at higher
energies involving orbitals away from the Fermi level, consistent
with the fact that the bulk boron is an indirect bandgap semiconductor,
with no optical absorption at the gap. In other words, optical absorption
features of bulk boron were already evident in smaller clusters. Given
the fact that aluminum has the same valence shell structure as boron,
it will be interesting to perform a similar set of calculations on
small aluminum clusters, in order to compare and contrast their photophysics.
Calculations along those directions are presently underway in our
group, and results will be communicated in future publications.

\section*{Acknowledgments}
R. S. thanks the Council of Scientific and Industrial Research (CSIR),
India, for a Junior Research Fellowship. 

\nonumsection{Appendix: Excited State CI Wavefunctions, Energies and Transition Moments}

\label{app:wavefunction} In the following tables we have given the
excitation energies (with respect to the ground state), the many body
wavefunction and the oscillator strengths of the excited states corresponding
to the peaks in the photoabsorption spectra of various isomers listed
in Fig. \ref{fig:geometries}, and discussed in section \ref{sec:results}. 

\begin{table*}
\tbl{\label{Tab:table_b2_lin}  Excitation energies, $E$, and many-particle wave functions of excited
states corresponding to the peaks in the linear absorption spectrum
of B$_{2}$ (\emph{cf}. Fig. \ref{fig:b2_lin_final}), along with
the oscillator strength of the transitions ($f_{12}=\frac{2}{3}\frac{m_{e}}{\hbar^{2}}(E_{2}-E_{1})\sum_{i}|\langle m|d_{i}|G\rangle|^{2}$),
where, $|m\rangle$ denotes the excited state in question, $|G\rangle$,
the ground state, and $d_{i}$ is the $i$-th Cartesian component
of the electric dipole operator. Longitudinal and transverse polarization
corresponds to the absorption due to light polarized along and perpendicular
to the molecular axis respectively. In the wave function, the bracketed
numbers are the CI coefficients of a given electronic configuration.
Symbols $H_{1}$,$H_{2}$ denote SOMOs discussed earlier, and $H$,
and $L$, denote HOMO and LUMO orbitals respectively. $HF$ denotes
the Hartree-Fock configuration. }
{\begin{tabular}{@{}ccccl@{}}
\toprule
Peak  & $E$ (eV)  & $f_{12}$  & Polarization & Wave Function \tabularnewline
\colrule
&  &  &  & \tabularnewline
GS$^a$
 &  &  &  & $|HF\rangle$ (0.8673)\tabularnewline
 &  &  &  & $|H-1\rightarrow L;H-1\rightarrow L\rangle$(0.2706)\tabularnewline
 &  &  &  & $|H-1\rightarrow L;H-1\rightarrow L+4\rangle$(0.1283)\tabularnewline
 &  &  &  & \tabularnewline
I  & 0.845 & 0.1187 & transverse & $|H_{2}\rightarrow L\rangle$(0.8742) \tabularnewline
 &  &  &  & $|H_{2}\rightarrow L+4\rangle$(0.2194) \tabularnewline
 &  &  &  & \tabularnewline
II  & 4.207 & 1.1449 & longitudinal & $|H-1\rightarrow L\rangle$(0.7599) \tabularnewline
 &  &  &  & $|H_{1}\rightarrow L+7\rangle$(0.2873) \tabularnewline
 &  &  &  & $|H_{2}\rightarrow L+7\rangle$(0.2873) \tabularnewline
 &  &  &  & \tabularnewline
III  & 4.914 & 0.4252 & transverse & $|H-1\rightarrow L;H-1\rightarrow L+2\rangle$(0.7553) \tabularnewline
 &  &  &  & $|H-2\rightarrow L+2\rangle$(0.3186) \tabularnewline
 &  &  &  & $|H-1\rightarrow L+7\rangle$(0.266) \tabularnewline
 &  &  &  & \tabularnewline
IV  & 6.97 & 4.3060 & transverse & $|H_{2}\rightarrow L+4\rangle$(0.5600) \tabularnewline
 &  &  &  & $|H_{2}\rightarrow L+5\rangle$(0.5150) \tabularnewline
 & 7.05 & 18.0923 & longitudinal & $|H-1\rightarrow L\rangle$(0.4462) \tabularnewline
 &  &  &  & $|H_{2}\rightarrow L+3\rangle$(0.3346) \tabularnewline
 &  &  &  & $|H_{1}\rightarrow L+3\rangle$(0.3346) \tabularnewline
 &  &  &  & $|H-1\rightarrow L+4\rangle$(0.2921) \tabularnewline
 &  &  &  & \tabularnewline
V & 7.973 & 0.6907 & longitudinal & $|H-1\rightarrow L;H_{2}\rightarrow L+2\rangle$(0.5613) \tabularnewline
 &  &  &  & $|H-1\rightarrow L;H_{1}\rightarrow L+2\rangle$(0.5613) \tabularnewline
 &  &  &  & $|H-1\rightarrow L+4\rangle$(0.2303) \tabularnewline
\botrule
\end{tabular} }
\begin{tabnote}
$^a${GS does not correspond to any peak, instead it corresponds to the
ground state wavefunction of the isomer.}
\end{tabnote}
\end{table*}

\FloatBarrier

\begin{table*}
\tbl{\label{Tab:table_b3_tri}  Excitation energies, $E$, and many-particle wave functions of excited
states corresponding to the peaks in the linear absorption spectrum
of triangular B$_{3}$ (\emph{cf}. Fig. \ref{fig:b3_tri_combined}),
along with oscillator strength of transition ($f_{12}=\frac{2}{3}\frac{m_{e}}{\hbar^{2}}(E_{2}-E_{1})\sum_{i}|\langle m|d_{i}|G\rangle|^{2}$),
where, $|m\rangle$ denotes the excited state in question, $|G\rangle$,
the ground state, and $d_{i}$ is the $i$-th Cartesian component
of the electric dipole operator. The polarization $\Vert$ corresponds
to the absorption due to light polarized in the plane of isomer, while
$\bot$ corresponds to the polarization perpendicular to that plane.
In the wave function, the bracketed numbers are the CI coefficients
of a given electronic configuration. Symbol $L$ and $H$ denote LUMO
and SOMO orbitals discussed earlier. $HF$ denotes the Hartree-Fock
configuration. }
{\begin{tabular}{@{}ccccl@{}}
\toprule
Peak  & $E$ (eV)  & $f_{12}$  & Polarization & Wave Function \tabularnewline
\colrule
 &  &  &  & \tabularnewline
GS$^a$
 &  &  &  & $|HF\rangle$ (0.8229)\tabularnewline
 &  &  &  & $|H-3\rightarrow L\rangle$(0.2162) \tabularnewline
 &  &  &  & $|H\rightarrow L\rangle$(0.2072) \tabularnewline
 &  &  &  & $|H-3\rightarrow H\rangle$(0.1544) \tabularnewline
 &  &  &  & \tabularnewline
I & 0.797 & 0.0912 & $\bot$ & $|H-1\rightarrow H\rangle$(0.7357) \tabularnewline
 &  &  &  & $|H-1\rightarrow L\rangle$(0.4693) \tabularnewline
 &  &  &  & $|H-1\rightarrow L+8\rangle$(0.1142) \tabularnewline
 &  &  &  & \tabularnewline
II & 1.710 & 0.0266 & $\Vert$ & $|H\rightarrow L+1\rangle$(0.4816) \tabularnewline
 &  &  &  & $|H-2\rightarrow H\rangle$(0.4315)\tabularnewline
 &  &  &  & $|H-2\rightarrow L\rangle$(0.3757)\tabularnewline
 &  &  &  & $|H-3\rightarrow L+1\rangle$(0.3709)\tabularnewline
 &  & 0.0282 & $\Vert$ & $|H\rightarrow L\rangle$(0.5041)\tabularnewline
 &  &  &  & $|H-3\rightarrow H\rangle$(0.4703)\tabularnewline
 &  &  &  & \tabularnewline
III & 2.840 & 0.5173 & $\Vert$ & $|H-2\rightarrow H\rangle$(0.5374) \tabularnewline
 &  &  &  & $|H\rightarrow L+1\rangle$(0.5199) \tabularnewline
 &  &  &  & $|H-2\rightarrow L\rangle$(0.3306) \tabularnewline
 &  &  &  & $|H-3\rightarrow L+1\rangle$(0.2216)\tabularnewline
 & 2.872 & 0.4741 & $\Vert$ & $|H\rightarrow L\rangle$(0.4906) \tabularnewline
 &  &  &  & $|H-3\rightarrow L\rangle$(0.4499) \tabularnewline
 &  &  &  & \tabularnewline
IV  & 5.710  & 0.3259 & $\Vert$ & $|H-2\rightarrow L\rangle$(0.4114) \tabularnewline
 &  &  &  & $|H-2\rightarrow H\rangle$(0.3474) \tabularnewline
 & 5.730 & 0.2053 & $\Vert$ & $|H-3\rightarrow L\rangle$(0.4202) \tabularnewline
 &  &  &  & $|H-2\rightarrow H;H-3\rightarrow L\rangle$(0.2050) \tabularnewline
 &  &  &  & \tabularnewline
V  & 5.988 & 0.5051 & $\Vert$ & $|H-3\rightarrow L+1\rangle$(0.3968)\tabularnewline
 &  &  &  & $|H-2\rightarrow L\rangle$(0.3057) \tabularnewline
 &  &  &  & $|H-2\rightarrow H\rangle$(0.1578) \tabularnewline
 & 6.02 & 0.9679 & $\Vert$ & $|H-3\rightarrow L\rangle$(0.2578) \tabularnewline
 &  &  &  & $|H\rightarrow L\rangle$(0.1908)\tabularnewline
 &  &  &  & \tabularnewline
VI  & 7.657 & 1.6393 & $\Vert$ & $|H\rightarrow L+2\rangle$(0.5863) \tabularnewline
 &  &  &  & $|H-3\rightarrow L+2\rangle$(0.2497)\tabularnewline
 &  &  &  & $|H\rightarrow L+14\rangle$(0.2431)\tabularnewline
 &  &  &  & $|H\rightarrow L+4\rangle$(0.2093) \tabularnewline
 & 7.697  & 1.0733 & $\Vert$ & $|H\rightarrow L+3\rangle$(0.2991) \tabularnewline
 &  &  &  & $|H-3\rightarrow L+1\rangle$(0.2648)\tabularnewline
 &  &  &  & $|H-2\rightarrow H\rangle$(0.1794) \tabularnewline
 &  &  &  & $|H-2\rightarrow L\rangle$(0.1521) \tabularnewline
 &  &  &  & \tabularnewline
VII & 7.893 & 2.1315 & $\bot$ & $|H\rightarrow L+5\rangle$(0.7611) \tabularnewline
 &  &  &  & $|H-3\rightarrow L+4\rangle$(0.3477) \tabularnewline
\botrule
\end{tabular}}
\begin{tabnote}
 $^a${GS does not correspond to any peak, instead it corresponds to the
ground state wavefunction of the isomer.}
\end{tabnote}
\end{table*}

\FloatBarrier

\begin{table*}
\tbl{\label{Tab:table_b3_lin} Excitation energies, $E$, and many-particle wave functions of excited
states corresponding to the peaks in the linear absorption spectrum
of linear B$_{3}$ (\emph{cf}. Fig. \ref{fig:b3_lin_final}), along
with oscillator strength of transition ($f_{12}=\frac{2}{3}\frac{m_{e}}{\hbar^{2}}(E_{2}-E_{1})\sum_{i}|\langle m|d_{i}|G\rangle|^{2}$),
where, $|m\rangle$ denotes the excited state in question, $|G\rangle$,
the ground state, and $d_{i}$ is the $i$-th Cartesian component
of the electric dipole operator. Longitudinal and transverse polarization
corresponds to the absorption due to light polarized along and perpendicular
to the molecular axis respectively. In the wave function, the bracketed
numbers are the CI coefficients of a given electronic configuration.
Symbols $H$ and $L$ denote HOMO and LUMO orbitals respectively,
and $H_{1}$ denotes SOMOs discussed earlier. $HF$ denotes the Hartree-Fock
configuration. }
{\begin{tabular}{@{}ccccl@{}}
\toprule
Peak  & $E$ (eV)  & $f_{12}$  & Polarization & Wave Function \tabularnewline
\colrule
&  &  &  & \tabularnewline
GS$^a$
 &  &  &  & $|HF\rangle$ (0.6650)\tabularnewline
 &  &  &  & $|H-1\rightarrow L;H-1\rightarrow L\rangle$(0.3286)\tabularnewline
 &  &  &  & $|H-1\rightarrow L;H-1\rightarrow L\rangle$(0.3277)\tabularnewline
 &  &  &  & $|H\rightarrow L;H-1\rightarrow L+1\rangle$(0.2158)\tabularnewline
 &  &  &  & $|H\rightarrow L;H-1\rightarrow L+1\rangle$(0.2157)\tabularnewline
 &  &  &  & \tabularnewline
I & 0.723 & 0.1520 & longitudinal & $|H\rightarrow L\rangle$(0.8222) \tabularnewline
 &  &  &  & \tabularnewline
II  & 2.707 & 0.1337 & transverse & $|H-1\rightarrow L\rangle$(0.5360) \tabularnewline
 &  &  &  & \tabularnewline
III & 4.338 & 4.5187 & longitudinal & $|H-1\rightarrow L;H-1\rightarrow H\rangle$(0.5826) \tabularnewline
 &  &  &  & $|H-1\rightarrow L;H-1\rightarrow H\rangle$(0.5826) \tabularnewline
 &  &  &  & \tabularnewline
IV  & 5.937 & 0.2295 & longitudinal & $|H-2\rightarrow L\rangle$(0.2497) \tabularnewline
 &  &  &  & \tabularnewline
V  & 7.359 & 8.4221 & longitudinal & $|H-1\rightarrow L+1\rangle$(0.3683) \tabularnewline
 &  &  &  & $|H-1\rightarrow L+1\rangle$(0.3683) \tabularnewline
 &  &  &  & $|H-3\rightarrow L;H\rightarrow L\rangle$(0.3301) \tabularnewline
 &  &  &  & $|H-3\rightarrow H\rangle$(0.2364) \tabularnewline
 &  &  &  & \tabularnewline
VI  & 7.731 & 2.9881 & longitudinal & $|H-2\rightarrow L\rangle$(0.3858) \tabularnewline
 &  &  &  & $|H\rightarrow L;H-1\rightarrow L+4\rangle$(0.2851) \tabularnewline
 &  &  &  & $|H\rightarrow L;H-1\rightarrow L+4\rangle$(0.2851)\tabularnewline
 & 7.786 & 2.3331 & transverse & $|H-1\rightarrow L;H-1\rightarrow L+4\rangle$(0.6438)\tabularnewline
 &  &  &  & $|H-1\rightarrow L;H-1\rightarrow L+4\rangle$(0.3905)\tabularnewline
\botrule
\end{tabular} }
\begin{tabnote}
$^a${GS does not correspond to any peak, instead it corresponds to the
ground state wavefunction of the isomer.}
\end{tabnote}
\end{table*}

\FloatBarrier

\begin{table*}
\tbl{\label{Tab:table_b4_rho} Excitation energies, $E$, and many-particle wave functions of excited
states corresponding to the peaks in the linear absorption spectrum
of rhombus B$_{4}$ (\emph{cf}. Fig. \ref{fig:b4_rho_combined}),
along with oscillator strength of transition ($f_{12}=\frac{2}{3}\frac{m_{e}}{\hbar^{2}}(E_{2}-E_{1})\sum_{i}|\langle m|d_{i}|G\rangle|^{2}$),
where, $|m\rangle$ denotes the excited state in question, $|G\rangle$,
the ground state, and $d_{i}$ is the $i$-th Cartesian component
of the electric dipole operator. The polarization \emph{in-plane}
corresponds to the absorption due to light polarized in the plane
of isomer. In the wave function, the bracketed numbers are the CI
coefficients of a given electronic configuration. Symbols $H$/$L$
denote HOMO/LUMO orbitals. $HF$ denotes the Hartree-Fock configuration. }
{\begin{tabular}{@{}ccccl@{}}
\toprule
Peak  & $E$ (eV) & $f_{12}$  & Polarization & Wave Function \tabularnewline
\colrule
&  &  &  & \tabularnewline
GS$^a$
 &  &  &  & $|HF\rangle$ (0.8787)\tabularnewline
 &  &  &  & $|H\rightarrow L;H\rightarrow L\rangle$(0.1147) \tabularnewline
 &  &  &  & \tabularnewline
I  & 4.159  & 3.5758 & in-plane & $|H\rightarrow L+2\rangle$(0.6566) \tabularnewline
 &  &  &  & $|H\rightarrow L+10\rangle$(0.3210) \tabularnewline
 &  &  &  & $|H-1\rightarrow L+6\rangle$(0.2773) \tabularnewline
 &  &  &  & $|H-1\rightarrow L+17\rangle$(0.1850)\tabularnewline
 &  &  &  & \tabularnewline
II  & 6.118  & 2.9911 & in-plane & $|H-1\rightarrow L+6\rangle$(0.5786) \tabularnewline
 &  &  &  & $|H-1\rightarrow L+17\rangle$(0.3285)\tabularnewline
 &  &  &  & $|H-3\rightarrow L\rangle$(0.2656) \tabularnewline
 &  &  &  & $|H-2\rightarrow L+11\rangle$(0.2544)\tabularnewline
 &  &  &  & $|H-2\rightarrow L+18\rangle$(0.2492)\tabularnewline
 &  &  &  & \tabularnewline
III  & 6.639  & 0.3735 & in-plane & $|H-1\rightarrow L+3\rangle$(0.4496)\tabularnewline
 &  &  &  & $|H-1\rightarrow L+13\rangle$(0.4485)\tabularnewline
 &  &  &  & $|H-4\rightarrow L\rangle$(0.4052) \tabularnewline
 &  &  &  & $|H\rightarrow L+11\rangle$(0.2766) \tabularnewline
 &  &  &  & $|H\rightarrow L+1\rangle$(0.2329) \tabularnewline
 &  &  &  & $|H\rightarrow L+18\rangle$(0.1905) \tabularnewline
 &  &  &  & \tabularnewline
IV  & 7.311  & 1.9267 & in-plane & $|H\rightarrow L+2\rangle$(0.3055) \tabularnewline
 &  &  &  & $|H-2\rightarrow L+11\rangle$(0.2892) \tabularnewline
 &  &  &  & $|H-3\rightarrow L\rangle$(0.2834) \tabularnewline
 &  &  &  & $|H-2\rightarrow L+18\rangle$(0.2487) \tabularnewline
 &  &  &  & $|H-2\rightarrow L+1\rangle$(0.2029) \tabularnewline
 &  &  &  & $|H-2\rightarrow L+5\rangle$(0.1509) \tabularnewline
 &  &  &  & \tabularnewline
V  & 7.842  & 3.0105 & in-plane & $|H\rightarrow L+2\rangle$(0.4233) \tabularnewline
 &  &  &  & $|H\rightarrow L+10\rangle$(0.3270) \tabularnewline
 &  &  &  & $|H-2\rightarrow L+11\rangle$(0.2049) \tabularnewline
 &  &  &  & $|H\rightarrow L+20\rangle$(0.1946) \tabularnewline
 &  &  &  & $|H-2\rightarrow L+1\rangle$(0.1594) \tabularnewline
 &  &  &  & $|H-2\rightarrow L+18\rangle$(0.1582) \tabularnewline
\botrule
\end{tabular} }
\begin{tabnote}
$^a${GS does not correspond to any peak, instead it corresponds to the
ground state wavefunction of the isomer.}
\end{tabnote}
\end{table*}

\FloatBarrier

\begin{table*}
\tbl{\label{Tab:table_b4_sqr} Excitation energies, $E$, and many-particle wave functions of excited
states corresponding to the peaks in the linear absorption spectrum
of square B$_{4}$ (\emph{cf}. Fig. \ref{fig:b4_sqr_combined}), along
with oscillator strength of transition ($f_{12}=\frac{2}{3}\frac{m_{e}}{\hbar^{2}}(E_{2}-E_{1})\sum_{i}|\langle m|d_{i}|G\rangle|^{2}$),
where, $|m\rangle$ denotes the excited state in question, $|G\rangle$,
the ground state, and $d_{i}$ is the $i$-th Cartesian component
of the electric dipole operator. The polarization \emph{in-plane}
corresponds to the absorption due to light polarized in the plane
of isomer. In the wave function, the bracketed numbers are the CI
coefficients of a given electronic configuration. Symbols $H$/$L$
denote HOMO/LUMO orbitals. $HF$ denotes the Hartree-Fock configuration. }
{ \begin{tabular}{@{}ccccl@{}}
\toprule
Peak  & $E$ (eV)  & $f_{12}$  & Polarization & Wave Function \tabularnewline
\colrule
&  &  &  & \tabularnewline
GS$^a$
 &  &  &  & $|HF\rangle$ (0.8682)\tabularnewline
 &  &  &  & $|H\rightarrow L;H\rightarrow L\rangle$(0.1765) \tabularnewline
 &  &  &  & $|H-2\rightarrow L;H-2\rightarrow L\rangle$(0.0920) \tabularnewline
 &  &  &  & \tabularnewline
I  & 4.879  & 4.3269 & in-plane & $|H\rightarrow L;H\rightarrow L+1\rangle$(0.2946) \tabularnewline
 &  &  &  & $|H\rightarrow L;H-3\rightarrow L\rangle$(0.1909) \tabularnewline
 &  &  &  & $|H-2\rightarrow L;H-3\rightarrow L\rangle$(0.1651) \tabularnewline
 &  &  &  & \tabularnewline
II  & 5.462 & 0.1792 & in-plane & $|H\rightarrow L;H\rightarrow L+1\rangle$(0.5893) \tabularnewline
 &  &  &  & $|H\rightarrow L;H\rightarrow L+6\rangle$(0.2477) \tabularnewline
 &  &  &  & \tabularnewline
III  & 6.418  & 0.2165 & in-plane & $|H\rightarrow L;H-3\rightarrow L\rangle$(0.5379) \tabularnewline
 &  &  &  & $|H\rightarrow L;H\rightarrow L+1\rangle$(0.2922) \tabularnewline
 &  &  &  & $|H-2\rightarrow L;H-3\rightarrow L\rangle$(0.1744) \tabularnewline
 &  &  &  & \tabularnewline
IV & 7.890 & 10.4628 & in-plane & $|H\rightarrow L+1;H-1\rightarrow L\rangle$(0.2351) \tabularnewline
\botrule
\end{tabular} }
\begin{tabnote}
 $^a${GS does not correspond to any peak, instead it corresponds to the
ground state wavefunction of the isomer.} 
\end{tabnote}
\end{table*}

\FloatBarrier

\begin{table*}
\tbl{\label{Tab:table_b4_lin} Excitation energies, $E$, and many-particle wave functions of excited
states corresponding to the peaks in the linear absorption spectrum
of linear B$_{4}$ (\emph{cf}. Fig. \ref{fig:b4_lin_final}), along
with oscillator strength of transition ($f_{12}=\frac{2}{3}\frac{m_{e}}{\hbar^{2}}(E_{2}-E_{1})\sum_{i}|\langle m|d_{i}|G\rangle|^{2}$),
where, $|m\rangle$ denotes the excited state in question, $|G\rangle$,
the ground state, and $d_{i}$ is the $i$-th Cartesian component
of the electric dipole operator. Longitudinal and transverse polarization
corresponds to the absorption due to light polarized along and perpendicular
to the molecular axis respectively. In the wave function, the bracketed
numbers are the CI coefficients of a given electronic configuration.
Symbols $H$/$L$ denote HOMO/LUMO orbitals. $HF$ denotes the Hartree-Fock
configuration. }
{\begin{tabular}{@{}ccccl@{}}
\toprule
Peak  & $E$ (eV)  & $f_{12}$  & Polarization & Wave Function \tabularnewline
\colrule
&  &  &  & \tabularnewline
GS$^a$
 &  &  &  & $|HF\rangle$ (0.5636)\tabularnewline
 &  &  &  & $|H\rightarrow L\rangle$(0.4737) \tabularnewline
 &  &  &  & $|H-1\rightarrow L;H\rightarrow L+1\rangle$(0.2291) \tabularnewline
 &  &  &  & $|H-1\rightarrow L;H\rightarrow L+1\rangle$(0.2289) \tabularnewline
 &  &  &  & \tabularnewline
I  & 5.363 & 0.7433 & longitudinal & $|H\rightarrow L;H-1\rightarrow L+9\rangle$(0.3005) \tabularnewline
 &  &  &  & $|H\rightarrow L;H-1\rightarrow L+9\rangle$(0.3005) \tabularnewline
 &  &  &  & $|H\rightarrow L;H-1\rightarrow L+8\rangle$(0.2702) \tabularnewline
 &  &  &  & $|H\rightarrow L;H-1\rightarrow L+8\rangle$(0.2702) \tabularnewline
 &  &  &  & $|H\rightarrow L;H-1\rightarrow L+4\rangle$(0.2439) \tabularnewline
 &  &  &  & $|H\rightarrow L;H-1\rightarrow L+4\rangle$(0.2439) \tabularnewline
 &  &  &  & \tabularnewline
II  & 5.947 & 31.8150 & longitudinal & $|H-1\rightarrow L+1\rangle$(0.2957)\tabularnewline
 &  &  &  & $|H-1\rightarrow L+1\rangle$(0.2957)\tabularnewline
 &  &  &  & \tabularnewline
III  & 7.352 & 14.0546 & longitudinal & $|H\rightarrow L;H-2\rightarrow L\rangle$(0.6477)\tabularnewline
 &  &  &  & $|H-3\rightarrow L\rangle$(0.2966)\tabularnewline
 &  &  &  & $|H-1\rightarrow L+1\rangle$(0.2091)\tabularnewline
 &  &  &  & $|H-1\rightarrow L+1\rangle$(0.2091)\tabularnewline
 & 7.380 & 7.7738 & transverse & $|H-1\rightarrow L+13\rangle$(0.5253)\tabularnewline
\botrule
\end{tabular} }
\begin{tabnote}
 $^a${GS does not correspond to any peak, instead it corresponds to the
ground state wavefunction of the isomer.}
\end{tabnote}
\end{table*}

\FloatBarrier

\begin{table*}
\tbl{\label{Tab:table_b4_tetra}  Excitation energies, $E$, and many-particle wave functions of excited
states corresponding to the peaks in the linear absorption spectrum
of distorted tetrahedron B$_{4}$ (\emph{cf}. Fig. \ref{fig:b4_tetra_combined}),
along with oscillator strength of transition ($f_{12}=\frac{2}{3}\frac{m_{e}}{\hbar^{2}}(E_{2}-E_{1})\sum_{i}|\langle m|d_{i}|G\rangle|^{2}$),
where, $|m\rangle$ denotes the excited state in question, $|G\rangle$,
the ground state, and $d_{i}$ is the $i$-th Cartesian component
of the electric dipole operator. The polarization $x$,$y$ and $z$
corresponds to the absorption due to light polarized along $x-$,$y-$
and $z-$ axis respectively. In the wave function, the bracketed numbers
are the CI coefficients of a given electronic configuration. Symbols
$H$/$L$ denote HOMO/LUMO orbitals. $HF$ denotes the Hartree-Fock
configuration. }
{\begin{tabular}{@{}ccccl@{}}
\toprule
Peak  & $E$ (eV) & $f_{12}$  & Polarization & Wave Function \tabularnewline
\colrule
 &  &  &  & \tabularnewline
GS$^a$
 &  &  &  & $|HF\rangle$ (0.6493)\tabularnewline
 &  &  &  & $|H\rightarrow L;H\rightarrow L+2\rangle$(0.4695) \tabularnewline
 &  &  &  & $|H\rightarrow L;H-3\rightarrow L+2\rangle$(0.1547) \tabularnewline
 &  &  &  & \tabularnewline
I  & 1.000 & 0.0517 & x & $|H\rightarrow L\rangle$(0.8127) \tabularnewline
 &  &  &  & $|H\rightarrow L;H-1\rightarrow L+2\rangle$(0.2364)\tabularnewline
 & 1.111 & 0.0714 & y & $|H\rightarrow L+2;H\rightarrow L+1\rangle$(0.5944) \tabularnewline
 &  &  &  & $|H\rightarrow L;H\rightarrow L+1\rangle$(0.5121) \tabularnewline
 &  &  &  & \tabularnewline
II  & 3.609 & 0.2771 & x & $|H\rightarrow L;H-1\rightarrow L+2\rangle$(0.5094) \tabularnewline
 &  &  &  & $|H\rightarrow L;H-1\rightarrow L\rangle$(0.4535)\tabularnewline
 &  &  &  & $|H\rightarrow L+2\rangle$(0.3480) \tabularnewline
 & 3.754 & 0.3957 & y & $|H\rightarrow L+1;H\rightarrow L+2\rangle$(0.2391)\tabularnewline
 &  &  &  & \tabularnewline
III  & 4.48 & 0.4818 & z & $|H-1\rightarrow L\rangle$(0.4418)\tabularnewline
 &  &  &  & \tabularnewline
IV  & 4.96 & 0.5841 & z & $|H-1\rightarrow L\rangle$(0.2678) \tabularnewline
 &  &  &  & $|H\rightarrow L;H-3\rightarrow L\rangle$(0.2227) \tabularnewline
 &  &  &  & \tabularnewline
V & 5.92 & 0.6954 & x & $|H-3\rightarrow L\rangle$(0.2191) \tabularnewline
 &  &  &  & $|H\rightarrow L+2;H-4\rightarrow L\rangle$(0.2191) \tabularnewline
 &  &  &  & \tabularnewline
VI & 6.15 & 0.6819 & y & $|H\rightarrow L;H-3\rightarrow L+1\rangle$(0.2526) \tabularnewline
 &  &  &  & \tabularnewline
VII & 6.508 & 0.1217 & z & $|H-1\rightarrow L+2\rangle$(0.7131) \tabularnewline
 &  &  &  & $|H-1\rightarrow L\rangle$(0.2104) \tabularnewline
 &  &  &  & $|H\rightarrow L+2;H\rightarrow L+2\rangle$(0.2030) \tabularnewline
 &  &  &  & \tabularnewline
VIII & 6.858 & 1.4336 & x & $|H\rightarrow L;H\rightarrow L+10\rangle$(0.6912) \tabularnewline
 &  &  &  & $|H\rightarrow L+2;H\rightarrow L+10\rangle$(0.3415) \tabularnewline
 &  &  &  & $|H\rightarrow L+18;H\rightarrow L\rangle$(0.2759) \tabularnewline
 & 6.99 & 1.1338 & y & $|H\rightarrow L;H\rightarrow L+5\rangle$(0.4169) \tabularnewline
 &  &  &  & $|H\rightarrow L+2;H\rightarrow L+5\rangle$(0.3770) \tabularnewline
 &  &  &  & $|H\rightarrow L+3;H-1\rightarrow L\rangle$(0.2403) \tabularnewline
 &  &  &  & $|H-1\rightarrow L+1\rangle$(0.2156) \tabularnewline
 &  &  &  & \tabularnewline
IX & 7.80 & 0.5231 & z & $|H-1\rightarrow L;H-1\rightarrow L+2\rangle$(0.3386) \tabularnewline
 &  &  &  & $|H\rightarrow L;H-4\rightarrow L+2\rangle$(0.3164) \tabularnewline
 &  &  &  & $|H\rightarrow L;H-4\rightarrow L+2\rangle$(0.2987) \tabularnewline
 &  &  &  & $|H-1\rightarrow L;H-1\rightarrow L\rangle$(0.2722) \tabularnewline
 &  &  &  & $|H\rightarrow L+1;H-1\rightarrow L+3\rangle$(0.2174) \tabularnewline
\botrule
\end{tabular} }
\begin{tabnote}
 $^a$ {GS does not correspond to any peak, instead it corresponds to the
ground state wavefunction of the isomer.}
\end{tabnote}
\end{table*}
\FloatBarrier

\begin{table*}
\tbl{\label{Tab:table_b5_pen} Excitation energies, $E$, and many-particle wave functions of excited
states corresponding to the peaks in the linear absorption spectrum
of pentagon B$_{5}$ (\emph{cf}. Fig. \ref{fig:b5_pen_combined}),
along with oscillator strength of transition ($f_{12}=\frac{2}{3}\frac{m_{e}}{\hbar^{2}}(E_{2}-E_{1})\sum_{i}|\langle m|d_{i}|G\rangle|^{2}$),
where, $|m\rangle$ denotes the excited state in question, $|G\rangle$,
the ground state, and $d_{i}$ is the $i$-th Cartesian component
of the electric dipole operator. The polarization \emph{in-plane}
corresponds to the absorption due to light polarized in the plane
of isomer. In the wave function, the bracketed numbers are the CI
coefficients of a given electronic configuration. Symbols $H$ and
$L$ denote HOMO and LUMO orbitals respectively. $HF$ denotes the
Hartree-Fock configuration. }
{\begin{tabular}{@{}ccccl@{}}
\toprule
Peak  & $E$ (eV) & $f_{12}$  & Polarization & Wave Function  \tabularnewline
\colrule
GS$^a$
 &  &  &  & $|HF\rangle$ (0.8541)\tabularnewline
 &  &  &  & $|H-1\rightarrow L\rangle$(0.1906) \tabularnewline
 &  &  &  & \tabularnewline
I & 1.394 & 0.1805 & $\bot$ to the plane & $|H\rightarrow L+1\rangle$(0.8536) \tabularnewline
 &  &  &  & $|H\rightarrow L+11\rangle$(0.1779) \tabularnewline
 &  &  &  & \tabularnewline
II & 1.818 & 0.0734 & in-plane & $|H\rightarrow L\rangle$(0.6524) \tabularnewline
 &  &  &  & $|H-1\rightarrow H\rangle$(0.4900) \tabularnewline
 &  &  &  & $|H\rightarrow L+3\rangle$(0.2218) \tabularnewline
 &  &  &  & \tabularnewline
III & 3.504  & 0.6466 & in-plane & $|H-2\rightarrow L+1\rangle$(0.5804) \tabularnewline
 &  &  &  & $|H-1\rightarrow H\rangle$(0.5131) \tabularnewline
 &  &  &  & $|H\rightarrow L\rangle$(0.3441) \tabularnewline
 &  &  &  & \tabularnewline
IV & 3.868 & 0.2839 & in-plane & $|H-2\rightarrow L+1\rangle$(0.7968) \tabularnewline
 &  &  &  & $|H-1\rightarrow H\rangle$(0.3416) \tabularnewline
 &  &  &  & \tabularnewline
V & 4.379 & 0.1885 & in-plane & $|H-3\rightarrow H\rangle$(0.8185) \tabularnewline
 &  &  &  & $|H-4\rightarrow L\rangle$(0.1900)\tabularnewline
 &  &  &  & \tabularnewline
VI & 5.378 & 1.2515 & in-plane & $|H-1\rightarrow L\rangle$(0.7309) \tabularnewline
 &  &  &  & $|H-4\rightarrow L\rangle$(0.3583) \tabularnewline
 &  &  &  & $|H-1\rightarrow L+3\rangle$(0.1994) \tabularnewline
 &  &  &  & $|H-3\rightarrow H\rangle$(0.1808) \tabularnewline
 & 5.576  & 5.0600 & in-plane & $|H-2\rightarrow L+1\rangle$(0.4415) \tabularnewline
 &  &  &  & $|H-3\rightarrow L\rangle$(0.4045) \tabularnewline
 &  &  &  & $|H-4\rightarrow H\rangle$(0.4000) \tabularnewline
 &  &  &  & $|H\rightarrow L\rangle$(0.2399) \tabularnewline
 &  &  &  & $|H-1\rightarrow L+4\rangle$(0.2039) \tabularnewline
 &  & Continued to Next Page &  & \tabularnewline
\colrule 
\end{tabular} }
\begin{tabnote}
 $^a$ {GS does not correspond to any peak, instead it corresponds to the
ground state wavefunction of the isomer.}
\end{tabnote}
\end{table*}

\begin{table*} 
\tbl{(\emph{Continued from last page}) Excitation energies, $E$, and many-particle wave functions of excited
states corresponding to the peaks in the linear absorption spectrum
of pentagon B$_{5}$ (\emph{cf}. Fig. \ref{fig:b5_pen_combined}),
along with oscillator strength of transition ($f_{12}=\frac{2}{3}\frac{m_{e}}{\hbar^{2}}(E_{2}-E_{1})\sum_{i}|\langle m|d_{i}|G\rangle|^{2}$),
where, $|m\rangle$ denotes the excited state in question, $|G\rangle$,
the ground state, and $d_{i}$ is the $i$-th Cartesian component
of the electric dipole operator. The polarization \emph{in-plane}
corresponds to the absorption due to light polarized in the plane
of isomer. In the wave function, the bracketed numbers are the CI
coefficients of a given electronic configuration. Symbols $H$ and
$L$ denote HOMO and LUMO orbitals respectively. $HF$ denotes the
Hartree-Fock configuration. }
{\begin{tabular}{@{}ccccl@{}}
\toprule
Peak  & $E$ (eV) & $f_{12}$  & Polarization & Wave Function  \tabularnewline
\colrule
VII & 6.305  & 4.9168 & in-plane & $|H-3\rightarrow L\rangle$(0.4073) \tabularnewline
 &  &  &  & $|H-1\rightarrow L+4\rangle$(0.3328) \tabularnewline
 &  &  &  & $|H-1\rightarrow L+9\rangle$(0.2952) \tabularnewline
 &  &  &  & $|H-2\rightarrow L+1\rangle$(0.2870) \tabularnewline
 &  &  &  & $|H-4\rightarrow L+4\rangle$(0.2772) \tabularnewline
 &  &  &  & $|H-4\rightarrow L+9\rangle$(0.2531) \tabularnewline
 & 6.528 & 0.7269 & in-plane & $|H-1\rightarrow L\rangle$(0.5596) \tabularnewline
 &  &  &  & $|H-4\rightarrow L\rangle$(0.3400) \tabularnewline
 &  &  &  & $|H-2\rightarrow L+2\rangle$(0.2761) \tabularnewline
 &  &  &  & $|H-3\rightarrow L+4\rangle$(0.2667) \tabularnewline
 &  &  &  & $|H-3\rightarrow L+9\rangle$(0.2598) \tabularnewline
 &  &  &  & \tabularnewline 
VIII & 7.161  & 1.9516 & in-plane & $|H-4\rightarrow L+4\rangle$(0.4305) \tabularnewline
 &  &  &  & $|H-4\rightarrow L+9\rangle$(0.3859) \tabularnewline
 &  &  &  & $|H-3\rightarrow L\rangle$(0.3776) \tabularnewline
 &  &  &  & $|H-1\rightarrow L+4\rangle$(0.2487) \tabularnewline
 & 7.283 & 0.5947 & in-plane & $|H\rightarrow L+6\rangle$(0.7102) \tabularnewline
 &  &  &  & $|H\rightarrow L+9\rangle$(0.3359)\tabularnewline
 &  &  &  & $|H-3\rightarrow L+4\rangle$(0.2091)\tabularnewline
 &  &  &  & $|H-3\rightarrow L+9\rangle$(0.1854)\tabularnewline
 &  &  &  & \tabularnewline
IX & 7.702 & 1.0247 & $\bot$ to the plane & $|H\rightarrow L+11\rangle$(0.7980) \tabularnewline
 &  &  &  & $|H\rightarrow L+14\rangle$(0.2692) \tabularnewline
 & 7.750  & 0.8764 & in-plane & $|H-1\rightarrow L+3\rangle$(0.8332) \tabularnewline
 &  &  &  & $|H-1\rightarrow L+19\rangle$(0.1555) \tabularnewline
\botrule
\end{tabular} }
\end{table*}

\FloatBarrier

\begin{table*}
\tbl{\label{Tab:table_b5_tet} Excitation energies, $E$, and many-particle wave functions of excited
states corresponding to the peaks in the linear absorption spectrum
of distorted triangular-bipyramid B$_{5}$ (\emph{cf}. Fig. \ref{fig:b5_tet_combined}),
along with oscillator strength of transition ($f_{12}=\frac{2}{3}\frac{m_{e}}{\hbar^{2}}(E_{2}-E_{1})\sum_{i}|\langle m|d_{i}|G\rangle|^{2}$),
where, $|m\rangle$ denotes the excited state in question, $|G\rangle$,
the ground state, and $d_{i}$ is the $i$-th Cartesian component
of the electric dipole operator. The polarization $x$,$y$ and $z$
corresponds to the absorption due to light polarized along $x-$,$y-$
and $z-$ axis respectively. In the wave function, the bracketed numbers
are the CI coefficients of a given electronic configuration. Symbols
$H$ and $L$ denote HOMO and LUMO orbitals respectively. $HF$ denotes
the Hartree-Fock configuration. }
{\begin{tabular}{@{}ccccl@{}}
\toprule
Peak  & $E$ (eV)  & $f_{12}$  & Polarization & Wave Function \tabularnewline
\colrule
 &  &  &  & \tabularnewline
GS$^a$
 &  &  &  & $|HF\rangle$ (0.8615)\tabularnewline
 &  &  &  & $|H-1\rightarrow H\rangle$(0.2165) \tabularnewline
 &  &  &  & $|H-1\rightarrow H;H-1\rightarrow L\rangle$(0.1371) \tabularnewline
 &  &  &  & $|H-2\rightarrow H;H-1\rightarrow L\rangle$(0.1143) \tabularnewline
 &  &  &  & \tabularnewline
I  & 1.774  & 0.4172 & z & $|H-1\rightarrow L+4\rangle$(0.8689) \tabularnewline
 &  &  &  & \tabularnewline
II  & 3.161  & 0.5817 & y & $|H-2\rightarrow H\rangle$(0.8734) \tabularnewline
 &  &  &  & $|H-1\rightarrow H;H-1\rightarrow L\rangle$(0.2303) \tabularnewline
 &  &  &  & \tabularnewline
III  & 4.274  & 0.3879 & z & $|H-1\rightarrow L+2\rangle$(0.8666) \tabularnewline
 &  &  &  & $|H-1\rightarrow L+6\rangle$(0.1775) \tabularnewline
 &  &  &  & \tabularnewline
IV  & 5.487  & 1.4856 & y & $|H-1\rightarrow H;H-1\rightarrow L\rangle$(0.7812) \tabularnewline
 &  &  &  & $|H-2\rightarrow H\rangle$(0.2336) \tabularnewline
 &  &  &  & $|H-2\rightarrow H;H-1\rightarrow L\rangle$(0.2332) \tabularnewline
 &  &  &  & \tabularnewline
V  & 6.408  & 0.8914 & z & $|H-4\rightarrow H\rangle$(0.7837) \tabularnewline
 &  &  &  & $|H-2\rightarrow L+2\rangle$(0.3357) \tabularnewline
 &  &  &  & $|H-2\rightarrow H;H-1\rightarrow L+1\rangle$(0.2274) \tabularnewline
 &  &  &  & \tabularnewline
VI  & 7.519  & 3.4471 & x & $|H-2\rightarrow H;H-1\rightarrow L+2\rangle$(0.8133) \tabularnewline
 &  &  &  & $|H-2\rightarrow H;H-1\rightarrow L+6\rangle$(0.2669) \tabularnewline
 &  &  &  & \tabularnewline
VII  & 7.744  & 0.1349 & x & $|H-1\rightarrow L;H-1\rightarrow L+1\rangle$(0.6105) \tabularnewline
 &  &  &  & $|H-1\rightarrow L+4\rangle$(0.4305) \tabularnewline
 &  &  &  & $|H-2\rightarrow L+1\rangle$(0.2271) \tabularnewline
\botrule
\end{tabular} }
\begin{tabnote}
 $^a$ {GS does not correspond to any peak, instead it corresponds to the
ground state wavefunction of the isomer.}
\end{tabnote}
\end{table*}

\FloatBarrier

\end{multicols}
\end{document}